\documentclass[smallextended]{svjour3}
\usepackage{amsmath}
\usepackage[doublespacing]{setspace}
\usepackage{graphicx}

\input{tcilatex}
\begin{document}

\begin{center}
\textbf{Towards a replicator dynamics model of age structured populations}%
\bigskip

K.~Argasinski* \\[0pt]
Institute of Mathematics of Polish Academy of Sciences\\[0pt]

ul. \'{S}niadeckich 8\\[0pt]

00-656 Warszawa\\[0pt]

Department of Mathematics, University of Sussex, \\[0pt]
Brighton BN1 9QH, UK.

\bigskip

\vspace{0.5cm} M.~Broom \\[0pt]
Department of Mathematics, City, University of London, \\[0pt]
Northampton Square, London EC1V 0HB, UK.

Mark.Broom.1@city.ac.uk\\[0pt]

\bigskip 

Journal of Mathematical Biology 2021 

doi.org/10.1007/s00285-021-01592-4
\end{center}

*corresponding author

The project is realized under grant Marie Curie grant\ PIEF-GA-2009-253845.

\newpage

\textbf{Abstract}

We present a new modelling framework combining replicator dynamics, the
standard model of frequency dependent selection, with an age-structured
population model. The new framework allows for the modelling of populations
consisting of competing strategies carried by individuals who change across
their life cycle. Firstly the discretization of the McKendrick von Foerster
model is derived. We show that the Euler--Lotka equation is satisfied when
the new model reaches a steady state (i.e. stable frequencies between the
age classes). This discretization consists of unit age classes where the
timescale is chosen so that only a fraction of individuals play a single
game round. This implies a linear dynamics and individuals not killed during
the round are moved to the next age class; linearity means that the system
is equivalent to a large Bernadelli-Lewis-Leslie matrix. Then we use the
methodology of multipopulation games to derive two, mutually equivalent
systems of equations. The first contains equations describing the evolution
of the strategy frequencies in the whole population, completed by subsystems
of equations describing the evolution of the age structure for each
strategy. The second contains equations describing the changes of the
general population's age structure, completed with subsystems of equations
describing the selection of the strategies within each age class. We then
present the obtained system of replicator dynamics in the form of the mixed
ODE-PDE system which is independent of the chosen timescale, and much
simpler. The obtained results are illustrated by the example of the sex
ratio model which shows that when different mortalities of the sexes are
assumed, the sex ratio of 0.5 is obtained but that Fisher's mechanism,
driven by the reproductive value of the different sexes, is not in
equilibrium. \bigskip

\textbf{Acknowledgements:} The project was realized under the grants Marie
Curie Actions PIEF-GA-2009-253845 by European Commision and
2013/08/S/NZ8/00821 FUGA 2 by the Polish National Science Centre. We want to
thank Jan Koz\l owski, John McNamara, Franjo Weissing for support for the
projects and Mats Gyllenberg, Reinhard B\"{u}rger, Ryszard Rudnicki, and
three anonymous reviewers for helpful suggestions. in memory Of Andrzej
Wieczorek.

\section{Introduction}

Among the most important approaches to the modelling of evolutionary
processes are life history optimization and evolutionary games. Classical
life history theory (Stearns 1992, Roff 1992) relies on optimization models,
where there are no interactions among individuals and no density dependence:

\emph{``Life history evolution usually ignores density and frequency
dependence. The justification is convenience, not logic, or realism''}
(Stearns 1992).

On the other hand, in classical game theoretic models there is no age or
stage structure. Payoffs describe the averaged lifetime activity of an
individual, which can be found for example in Cressman (1992):

\emph{\textquotedblleft ...an individual's strategy is fixed over its
lifetime or, alternatively, the life history of an individual is its
strategy.\textquotedblright }

Thus the synthesis of these perspectives can be very fruitful for
theoretical insight (McNamara 2013). Methods used in life history
optimization are closely related to classical demographic methods such as
Bernadelli-Lewis-Leslie matrices (Caswell 2001). However, how to construct a
general description of the relationships between demographic structure and
population dynamics is still an unsolved problem (Caswell 2011). More
precise than matrix models are continuous approaches arising from Lotka's
renewal equation (Lotka 1911, Diekmann et al. 2020a, 2020b) and the
McKendrick von Foerster model (McKendrick 1926). The combination of
demography with a game theoretic perspective focused on frequency dependent
selection, advocated by McNamara (2013), can be very useful since
demographers are interested in the patterns produced by heterogeneity in the
populations (Vaupel et al. 1979, Vaupel and Yashin 1983, Hougaard
1984,Vaupel and Yashin 1985). The game theoretic structure can explain the
mechanisms shaping those patterns. The first papers combining both
approaches are Garay et. al (2016) devoted to the particular biological
problem of sib cannibalism, Li et al. (2015) and Lessard and Soares (2017)
containing the approach incorporating age structure into a matrix game.
These results show that after introduction of the age structure, matrix
notation becomes very complicated and makes analysis difficult even in the
case of two competing strategies and few age classes. In addition, previous
works do not study the interplay between game dynamics and demographic
structure in detail, assuming a fixed demographic structure. However, the
game interactions described by demographic payoffs should affect the
demographic structures of subpopulations of carriers of different
strategies. In addition Li et al (2015) assumes that payoffs are described
by a standard payoff matrix, thus the same actions performed in different
ages/stages will generate the same payoffs. However, we can expect that
outcomes of individual actions may vary for different ages due to different
experiences and physical condition of the playing individuals.

Another problem is that game theoretic models operate in abstract terms of
costs and benefits measured in units of fitness mostly without deeper
insight into their meaning or interpretation. This problem was analyzed in
Argasinski and Broom (2012) where relationships between classical demography
and evolutionary games are described in detail. This approach was later
clarified in Argasinski and Broom (2018a, 2018b) by definition of the vital
rates (birth and death rates) as the product of the interaction rates,
describing the distribution of interactions (game rounds) in time and
demographic game payoffs describing the number of offspring and the
probability of death during a single interaction. The main conclusion there
is that instead of excess above average fitness, models should be described
explicitly by mortality and fertility, which are basic opposite forces
shaping population dynamics (Doebeli et al 2017). These results are
significant progress in ecological realism, emphasizing the role of
background mortality and fertility or the turnover of individuals
(Argasinski and Koz\l owski, 2008). However, that approach is still very
primitive. Mortality is described as an exponential decay of the population,
which implies that the length of an individual's lifetime is potentially
unbounded, and there is no aging and no age specific payoffs. The goal of
this paper is to fill this gap and develop a mathematical structure
combining selection of individual strategies with an age structured
population which will allow us to overcome the problems arising from
increasing complexity of the models shown in Li et al. (2015) and
simplifications ignoring the age dependence of payoffs resulting from
certain actions and feedbacks driving the interplay between game dynamics
and demography (the fixed age structure assumption). For practical reasons
we will develop a high dimensional ODE system consisting of relatively
simple equations, which can be generated by a simple loop and solved in
every popular numerical platform.

\subsection{The classical approach to evolutionary games and replicator
dynamics}

In the following subsections, we describe the state of the art in relation
to our problem. A list of existing (and indeed new, see later) parameters
are described in Table 1. Traditionally, in evolutionary game theory the
payoff obtained by the $j$th strategy is proportional to its Malthusian
growth rate $r_{j}$ and the dynamics of selection of strategies is described
by the \emph{replicator dynamics} (Maynard Smith 1982, Cressman 1992,
Hofbauer and Sigmund 1988 and 1998, Weibull 1995, Nowak 2006, Broom and
Rychtar 2013, McNamara and Leimar 2020)). We can derive this by rescaling
the Malthusian equations for competing strategies $\dot{n}_{j}=n_{j}r_{j}$
to relative frequencies $q_{j}=n_{j}/n$ (where $n=\sum_{j=1}^{w}n_{j}$ and $w
$ is the number of strategies), which leads to 
\begin{equation}
\dot{q}_{j}=q_{j}(r_{j}-\bar{r})  \label{REP}
\end{equation}%
where $\bar{r}=\sum_{j=1}^{w}q_{j}r_{j}$ is the average payoff in the
population. However, instead of the Malthusian parameter describing the
payoff we can explicitly consider the individual fertility $f_{j}$ and
mortality $d_{j}$ of a $j$-strategist. The explicit distinction between
fertility and mortality was proposed also by Doebeli et al (2017) as the
cornerstone of a mechanistic model of natural selection. Note that in real
life organisms are involved in different types of interactions with others
or elements of the environment. Game theoretic models are focused on the
outcomes of the particular interactions (such as fights as in the Hawk Dove
game) responsible for selection of the analyzed trait or type of behaviour.
These can be described by average demographic outcomes per interaction $f_{j}
$ and $d_{j}$ and these focal interactions will occur at the rate $\tau _{f}$%
. Other interactions, not related to the analyzed trait, can be described by
average fertility $f_{b}$ and mortality $d_{b}$, occurring at rate $\tau _{b}
$.

Products of interaction rates and demographic payoffs will constitute the
respective \textit{vital rates}: game fertility rate $\tau _{f}f_{j}$ and
mortality rate $\tau _{f}d_{j}$, background fertility rate $\tau _{b}f_{b}$
and mortality rate $\tau _{b}d_{b}$. Later the focal game interaction rate $%
\tau _{f}$ can be set to $1$ by timescale adjustment and the background
fertility and mortality rates become $\Phi =\tau _{b}f_{b}/\tau _{f}$ and $%
\Psi =\tau _{b}d_{b}/\tau _{f}$. In addition we can add density dependent
juvenile recruitment survival (Argasinski and Koz\l owski 2008, Argasinski
and Broom 2012, 2018a, 2018b). To do this we should multiply fertilities by
the logistic suppression coefficient $(1-n/K)$ (where the carrying capacity $%
K$ is interpreted as the maximal environmental load, Hui 2006). Since
fertilities but not mortalities are so scaled, the turnover of generations
will not be suppressed at the equilibrium as it is in the classical logistic
model (which leads to an immortal and childless population at equilibrium $K$%
). This gives the following variant of the replicator equations: 
\begin{eqnarray}
\dot{q}_{j} &=&q_{j}((f_{j}-\bar{f})\left( 1-\frac{n}{K}\right) -(d_{j}-\bar{%
d})),  \label{QWD} \\
\dot{n} &=&n(\left[ \bar{f}+\Phi \right] \left( 1-\frac{n}{K}\right) -\bar{d}%
-\Psi ),
\end{eqnarray}%
where $\bar{f}=\sum_{j=1}^{w}q_{j}f_{j}$ and $\bar{d}%
=\sum_{j=1}^{w}q_{j}d_{j}$, the details of which appear in Argasinski and
Broom (2012, 2018a, 2018b).

It was shown (Argasinski 2006) that every single population system described
by the replicator equations (\ref{REP}) can be divided into the product of
subsystems describing the dynamics in arbitrary chosen disjoint
subpopulations (described by a frequencies $q_{j}^{i}=n_{j}^{i}/n_{j}$,
where $n_{j}=\sum_{i}n_{j}^{i}$, for the $j$-th subpopulation) and an
additional system describing the dynamics of proportions between those
subpopulations $p_{j}=n_{j}/\sum_{z}n_{z}$. This is useful when indiviuals
differ not only by strategies but also by another second trait such as sex,
age or developmental stage. Then, for example, we can decompose the
population into subpopulations of carriers of the same strategy and describe
the dynamics of the second trait among them. Then the dynamics in each
subpopulation will have the form (\ref{REP}) and will depend on the excess
of the strategy payoff from the average payoff in this subpopulation.
Therefore, the same operation can be carried out for equations (\ref{QWD}),
and we obtain the system: 
\begin{eqnarray}
\dot{q}_{j}^{i} &=&q_{j}^{i}\left( \left( f_{j}^{i}-\bar{f}_{j}\right)
\left( 1-\frac{n}{K}\right) -(d_{j}^{i}-\bar{d}_{j})\right) ,  \label{INTRA}
\\
\dot{p}_{j} &=&p_{j}\left( \left( \bar{f}_{j}-\bar{f}\right) \left( 1-\frac{n%
}{K}\right) -(\bar{d}_{j}-\bar{d})\right) ,  \label{INTER} \\
\dot{n} &=&n(\left[ \bar{f}+\Phi \right] \left( 1-\frac{n}{K}\right) -\bar{d}%
-\Psi ),
\end{eqnarray}%
where $f_{j}^{i}$ and $d_{j}^{i}$ are the fertility and mortality,
respectively, of the $i$- th type (such as age or sex) in the subpopulation
of the $j$-th strategy carriers, $\bar{f}_{j}=%
\sum_{i=1}^{w}q_{j}^{i}f_{j}^{i}$ and $\bar{d}_{j}=%
\sum_{i=1}^{w}q_{j}^{i}d_{j}^{i}$ are the mean fertility and mortality,
respectively, in the subpopulation of the $j$-th strategy carriers and $\bar{%
f}$ and $\bar{d}$ are the respective values in the global population.

Note that we can decompose the initial population with respect to the second
trait and describe the dynamics of strategic composition among individuals
in the same age or sex class. Then the equations will describe variables $%
q_{j}^{i}$, $p_{j}$ and $n$.

\subsection{The classical approach to the modelling of age structured
populations}

Now we focus on age structured models (age classes will be indexed by
superscripts). The classical approach to the modelling of age structured
populations is related to Bernadelli-Lewis-Leslie matrices (Bernadelli 1941,
Lewis 1942, Leslie 1945, Charlesworth 1994, Caswell 2001), following the
matrix equation:

\begin{equation}
\left[ 
\begin{array}{c}
n^{0} \\ 
n^{1} \\ 
... \\ 
n^{m}%
\end{array}%
\right] _{t+1}=\left[ 
\begin{array}{cccc}
f^{0} & f^{1} & ... & f^{m} \\ 
s^{0} & 0 & 0 & 0 \\ 
0 & ... & 0 & 0 \\ 
0 & 0 & s^{m-1} & 0%
\end{array}%
\right] \left[ 
\begin{array}{c}
n^{0} \\ 
n^{1} \\ 
... \\ 
n^{m}%
\end{array}%
\right] _{t},  \label{LeslieMatrix}
\end{equation}%
where there are $m+1$ age classes, $n^{i}$ is the size of the $i$th age
class and $f^{i}$ is fertility and $s^{i}$ is survival, respectively, in
this class. Thus $n^{0}(t+1)=\sum_{i}n^{i}(t)f^{i}$ and the transition
between subsequent age classes is $n^{i}(t+1)=s^{i-1}n^{i-1}(t)$. When the
time unit equals the time step between age classes the above system is a
good model of age structure. This age-structured growth model suggests a
steady-state, or stable, age-structure and growth rate. The growth rate can
be calculated from the characteristic polynomial of the
Bernadelli-Lewis-Leslie Matrix called the Euler-Lotka equation\ (Caswell
2001): 
\begin{equation}
f^{0}+\sum\limits_{i=1}^{m}e^{-ir}f^{i}\prod\limits_{z=0}^{i-1}s^{z}=1,
\end{equation}%
where $r$ is the intrinsic growth rate of the population and\ $%
\prod\limits_{z=0}^{i-1}s^{z}$\ describes survival to age $i$. We note here
that in reality $r$ will not be an independent parameter, and moreover will
change in time as the distributions of the sizes of age classes change. An
equilibrium distribution over the age classes in turn will allow us to
define $r$ in terms of the other model parameters. A simple ODE
generalization of this system with continuous time but discrete age
structure can be obtained by application of the delayed differential
equations (Caswell 2001) where survival rates may describe aggregated
exponential survival between respective age classes (Diekmann et al 2017).
However this approach may not work if the mortality function depends on the
actual population state (as in game theory). Here the mortality rate may be
unknown since it will depend on the trajectory of the dynamics during the
age class. Then we can consider the continuous time limit of an infinite
number of infinitely small age classes where population structure becomes a
function $n(t,l)$ of time $t$ and continuous age $l$ describing a moment in
the lifetime of an individual. Then we can imagine the Taylor expansion
analogous to the transition equation describing a small time step $dt$
leading to ageing $dl$ 
\begin{equation}
n(t+dt,l+dl)=n(t,l)+\frac{\partial n}{\partial t}dt+\frac{\partial n}{%
\partial l}dl=s(l)n(t,l)=\left( 1-\tau d(l)dt\right) n(t,l),
\label{McTaylor}
\end{equation}%
where $\tau d(l)$ is the continuous time mortality rate (at age $l$)
similarly to the game models but without the distinction between the focal
game and the background interactions. Since $dl=dt$ we obtain the McKendrick
von Foerster equation 
\begin{equation}
\frac{\partial n(t,l)}{\partial t}+\frac{\partial n(t,l)}{\partial l}=-\tau
d(l)n(t,l),  \label{McKen}
\end{equation}%
which should be completed by boundary conditions $n(t,0)=\int_{0}^{\infty
}n(t,l)\tau f(l)dl$ and initial age distribution $n(0,l)$.

\section{The paper structure}

In this paper we derive the discretization of the McKendrick von Foerster
model allowing for the derivation of frequency dependent models. This is
motivated by the fact that the discretized approach can be easily
numerically solved by basic ODE solvers from popular numerical platforms.
Thus the developed methodology does not need advanced knowledge in numerical
analysis. Another advantage is that it will be compatible with standard game
theoretic notation based on matrix games. Using the derived discretization
we build two approaches to modelling selection among competing strategies
with life cycles in an asexual population. One is focused on the impact of
age structures of strategies on selection, while the second shows the impact
of selection dynamics on the age structure of the whole population. The
models obtained are generalized to mixed PDE-ODE models with continuous
non-discretized age structures to outline the direction of future
development. This framework is illustrated by a sex ratio example combining
the two approaches, allowing us to model the sexually reproducing
population. Our intention is to build a simple ready to use modelling
methodology which can be extended in the future. However, we believe that
even after clarification of the PDE based approach and development of simple
solvers of coupled integro-differential PDE-ODE systems, our approach will
still be useful for practical reasons arising from the simplicity of the
methods based on matrix payoffs, which are much simpler to derive than
continuous payoff functions. Thus it can be, for example, easily used for
building initial toy models.

\section{Results}

\subsection{Presenting the McKendrick von Foerster model as a system of ODE's%
}

In this section we will build the submodel describing the age structure
dynamics of a subpopulation of carriers of some strategy competing with
other strategies. Demographic vital rates will be outcomes of interactions
between carriers of different strategies, interpreted as rounds of
evolutionary games as in Argasinski and Broom (2018a). Thus as in replicator
dynamics models we have the state of the population described by strategy
frequencies $p_{j}$ but for each strategy subpopulation we have a respective
age structure described by parameters $a_{j}^{i}=n_{j}^{i}/n_{j}$\
(frequencies of individuals of age $i$\ among $j$-th strategy carriers).
Demographic payoffs determining the vital rates will depend not only on the
strategy frequencies $p=[p_{1},\ldots ,p_{w}]$ as in the classical
replicator models but also on the age of the opponents, thus the set of
vectors of the age structures for all strategies\ $a=[a_{1},\ldots ,a_{w}]$\
(where $a_{j}=[a_{j}^{0},\ldots ,a_{j}^{m}]$\ describes the age structure of
the $j$-th strategy carriers subpopulation) should be another argument of
the payoff functions.

A major technical difference between the McKendrick von Foerster model and
replicator dynamics is that the first is a PDE (or system of PDE's as for
example in Rudnicki and Mackey, 1993) and the second is a system of ODE's.
The simple combination of both approaches will lead to a mathematically
elegant but technically intractable system due to the lack of a general
theory for mixed PDE-ODE systems and software for numerical computation.
This methodology should be developed in the future, however before that, we
need a useful approach based on existing solutions. To solve this problem we
can approximate the continuous system by a large number of ODE's describing
unit interval age classes consisting of all individuals of age from $a$ to $%
a+1$. The discrete structure will allow us to use standard matrix payoff
functions. The chosen time unit should be as long as possible to reduce the
number of equations. Since we want to model frequency dependent selection,
the mortality and fertility payoffs will depend on the trajectory of the
population state. Therefore we cannot use simplified delayed differential
equations since we do not know the trajectories during the time delay
interval. Instead we can assume that the unit of a timescale described by
interaction rate $\tau $ is short enough that the changes of the population
state are small enough with respect to the population size (e.g. 50 births
in a population of 30000), that the resulting changes of frequency dependent
birth and death rates will be negligible. 

Following Appendix A we see that equation (\ref{McKen}) can be discretized
and approximated by the replicator dynamics\ (see Fig. 1 for the intuitive
presentation of the discretization scheme for frequency dependent vital
rates). In particular for the $j$th strategy we describe the system in
frequencies $a_{j}^{i}=n_{j}^{i}/\sum_{z=0}^{m}n_{j}^{z}$ and a scaling
parameter $n$. Assume that $\tilde{r}_{j}^{i}(t)=f_{j}^{i}(p(t),a(t))\left(
1-\dfrac{n}{K}\right) -d_{j}^{i}(p(t),a(t))$ is the game payoff component of
the growth rate (then $r_{j}^{i}(t)=\tau \tilde{r}_{j}^{i}(t)$) and $\tilde{r%
}_{j}(t)=\sum_{j}a_{j}^{i}\tilde{r}_{j}^{i}(t)$ is the respective averaged
value. If the growth rates $\tau \tilde{r}_{j}(t)$ are nearly constant, then
for the chosen timescale described by interaction rate $\tau $ changes of
the strategy frequencies during a single time unit are $\Delta p_{j}=\dfrac{%
\tau }{(1+\tau \tilde{r}(t))}p_{j}(t)\left( \tilde{r}_{j}(t)-\tilde{r}%
(t)\right) $ (where $\tilde{r}(t)=\sum_{i}p_{j}\tilde{r}_{j}(t)$), thus they
are sublinear. Here $\tau $ should be as big as possible to minimize the
number of equations, but small enough that payoff function arguments $\Delta
p_{j}$ (and similarly others) should change their values only slightly (i.e. 
$\Delta f_{j}^{i}=f_{j}^{i}(p(t)+\Delta p)-f_{j}^{i}(p(t))$ and $\Delta
d_{j}^{i}=d_{j}^{i}(p(t)+\Delta p)-d_{j}^{i}(p(t))$ are small enough, but
not necessarily infinitesimal), so that the resulting changes of $\tau
\Delta f_{j}^{i}$ and $\tau \Delta d_{j}^{i}$ are negligible. Then the
discretization is acceptable and we obtain: 
\begin{eqnarray}
\dot{a}_{j}^{i} &=&a_{j}^{i-1}s_{j}^{i-1}(p,a)-a_{j}^{i}\left( \bar{r}%
_{j}(p,a,n)+1\right) \hspace{0.6cm}i=1,...,m,  \label{agefreq} \\
\dot{n}_{j} &=&n_{j}\bar{r}_{j}(p,a,n),  \label{agescale}
\end{eqnarray}%
where $a_{j}^{0}=1-\sum_{i=1}^{m}a_{j}^{i}$ and the Malthusian parameter
describing the growth of the $j$th strategy is 
\begin{equation}
\bar{r}_{j}(p,a,n)=\sum_{i=0}^{m}a_{j}^{i}\left( \tau f_{j}^{i}(p,a)\left( 1-%
\dfrac{n}{K}\right) +s_{j}^{i}(p,a)\right) -1.  \label{r}
\end{equation}%
It is important that age class survival $s_{j}^{i}(p,a)=(1-\tau
d_{j}^{i}(p,a))$ describes aggregated outcomes of the game rounds occurring
during a time unit. Therefore it is distinct from the survival probability
of a single round $1-d_{j}^{i}(p,a)$ which should be used in trade-off
functions when only survivors of the game round can reproduce (Argasinski
and Broom 2012, 2018a,b), leading to fertility $(1-d_{j}^{i}(p,a))f_{j}^{i}$%
. In addition, due to nearly linear behaviour within a single time unit the
system (\ref{agefreq},\ref{agescale}) is equivalent to the large Leslie
matrix (\ref{LeslieMatrix}) with survival $s_{j}^{i}(p,a)=(1-\tau
d_{j}^{i}(p,a))$ and then parameter $\tau $ describes the fraction of
individuals that played the single game round. \ Parameter $\tau $\ always
acts as the multiplier of game payoffs $f_{j}^{i}$\ and $d_{j}^{i}$\ (thus
the resulting survival rate is $1-\tau d_{j}^{i}$).\ Since in the next
sections we will focus on the derivation of the dynamics, where the
structure of the vital rates is not so important, for simplicity\ we can
incorporate the interaction rate $\tau $ into the birth and death vital
rates and skip it in the notation. Therefore, below, $\tau $\ will be hidden
inside functions $f_{j}^{i}$ and $s_{j}^{i}$\ which will be interpreted as
the vital rates.

Assume the absence of density dependence. Since the r.h.s. of our system (%
\ref{agefreq}) is the negative function of $a_{j}^{i}$, the following
attracting nullcline manifold exists (for constant mortalities $s_{j}$ this
is an attracting steady state): 
\begin{equation}
\hat{a}_{j}^{i}=\dfrac{\hat{a}_{j}^{0}\prod\limits_{z=0}^{i-1}s_{j}^{z}(p,a)%
}{\left( \bar{r}_{j}(p.a)+1\right) ^{i}}=\dfrac{\hat{a}_{j}^{0}\prod%
\limits_{z=0}^{i-1}s_{j}^{z}(p,a)}{\left( \sum_{z=0}^{m}\hat{a}%
_{j}^{z}\left( f_{j}^{z}(p,a)+s_{j}^{z}(p,a)\right) \right) ^{i}}.
\label{stableage}
\end{equation}%
Note that $\hat{a}_{j}^{0}$ will satisfy the general form for $\hat{a}%
_{j}^{i}$ in equation (\ref{stableage}). In addition the Euler-Lotka
equation is satisfied (for a derivation and proof, see Appendix B). In the
density dependent case, the age structure attractor (\ref{stableage}) will
change with the growth of the population.\ Now we can use the derived
submodel for derivation of the full model.

\vspace{0.1cm}

FIGURE 1 HERE

\subsection{The extension to multipopulation replicator dynamics}

Now we can incorporate the above model into a multipopulation evolutionary
game (Argasinski 2006). Recall that we have $w$ strategies and $m+1$ age
classes indexed from $0$ to $m$. Assume that $p$ describes the strategy
(phenotype) fraction and $a$ describes the frequency of the age class. As
before, $f_{j}^{i}$ and $s_{j}^{i}$ describe, respectively, the fertility
and survival of the $j$-strategist in age class $i$. Two perspectives are
possible (see Fig. 2):

a) Firstly we consider the impact of the age structure in sub-populations
strategically homogenous on selection of the strategies, denoted as system $%
S_{a}$. This can be described by coordinates:


$a_{j}^{0},...,a_{j}^{m}$ \ \ \ \ \ for \ $j=1,...,w$ \ \ \ the age
structure of the $j$-strategists

$p_{1},....,p_{w}$ \ \ \ \ \ \ \ the strategy frequencies in the whole
population,

where $a_{j}^{i}=n_{j}^{i}/\sum_{z}n_{j}^{z}$ and $p_{j}=\sum_{z}n_{j}^{z}/n$%
.

b) Secondly we consider how selection within each age class affects the
overall age structure, denoted as system $S_{b}$. It can be described by
coordinates:


$p_{1}^{i},....,p_{w}^{i}$ \ \ \ \ \ for $i=0,...,m$ \ \ \ strategy
frequencies in age class $i$


$a^{0},...,a^{m}$ \ \ \ \ the age structure of the population,

where $p_{j}^{i}=n_{j}^{i}/\sum_{z}n_{z}^{i}$ and $a^{i}=\sum_{z}n_{z}^{i}/n$%
.

\vspace{0.1cm}

Thus in both cases we will have a core system describing the whole
population (strategic composition in $S_{a}$ and age structure in $S_{b}$)
completed by the respective subsystems describing the age structure of the
subpopulation of strategy carriers (for $S_{a}$) or the strategic age class
composition (for $S_{b}$).

FIGURE 2 HERE

Now we describe the transition of coordinates between the formulations.
First we define the auxiliary canonical coordinates without subclasses: 
\begin{equation}
q_{j}^{i}=a^{i}p_{j}^{i}=p_{j}a_{j}^{i}.  \label{q}
\end{equation}

Now following Argasinski (2006) we define transitions between systems: 
\newline
$S_{a}$ to $S_{b}$: 
\begin{eqnarray}
p^{i} &=&\left[ p_{1}^{i},...,p_{w}^{i}\right] =\left[ \dfrac{a_{1}^{i}p_{1}%
}{\sum_{j=1}^{w}a_{j}^{i}p_{j}},...,\dfrac{a_{w}^{i}p_{w}}{%
\sum_{j=1}^{w}a_{j}^{i}p_{j}}\right] ,  \label{atob1} \\
a &=&\left[ a^{0},...,a^{m}\right] =\left[ \sum_{j=1}^{w}a_{j}^{0}p_{j},...,%
\sum_{j=1}^{w}a_{j}^{m}p_{j}\right] ,  \label{atob2}
\end{eqnarray}%
and $S_{b}$ to $S_{a}$: 
\begin{eqnarray}
a_{j} &=&\left[ a_{j}^{0},...,a_{j}^{m}\right] =\left[ \dfrac{a^{0}p_{j}^{0}%
}{\sum_{i=0}^{m}a^{i}p_{j}^{i}},...,\dfrac{a^{m}p_{j}^{m}}{%
\sum_{i=0}^{m}a^{i}p_{j}^{i}}\right] ,  \label{btoa1} \\
p &=&\left[ p_{1},...,p_{w}\right] =\left[ \sum_{i=0}^{m}a^{i}p_{1}^{i},...,%
\sum_{i=0}^{m}a^{i}p_{w}^{i}\right] .  \label{btoa2}
\end{eqnarray}

Now let us derive systems of equations operating in both coordinate systems.
In the following we use the within group averaging terms: \newline
$\bar{f}_{j}=\sum_{i=0}^{m}a_{j}^{i}f_{j}^{i}$, $\bar{s}_{j}=%
\sum_{i=0}^{m}a_{j}^{i}s_{j}^{i}$, $\bar{s}^{i}=%
\sum_{j=1}^{w}p_{j}^{i}s_{j}^{i}$, $\ \bar{f}^{i}=%
\sum_{j=1}^{w}p_{j}^{i}f_{j}^{i}$. We also use two global averages, which
can each be written in two ways: $\bar{f}=\sum_{j=1}^{w}p_{j}\bar{f}%
_{j}=\sum_{i=0}^{m}a^{i}\bar{f}^{i}$ and $\bar{s}=\sum_{j=1}^{w}p_{j}\bar{s}%
_{j}=\sum_{i=0}^{m}a^{i}\bar{s}^{i}$. For system $S_{a}$ we have the
following system of differential equations (see Appendix C): 
\begin{eqnarray}
\dot{a}_{j}^{i} &=&a_{j}^{i-1}s_{j}^{i-1}-a_{j}^{i}\left( \bar{f}_{j}\left(
1-\frac{n}{K}\right) +\bar{s}_{j}\right)  \label{sysa1} \\
\dot{p}_{j} &=&p_{j}\left( \left( \bar{f}_{j}-\bar{f}\right) \left( 1-\frac{n%
}{K}\right) +\left( \bar{s}_{j}-\bar{s}\right) \right)  \label{sysa2} \\
\dot{n} &=&n\left( \bar{f}\left( 1-\frac{n}{K}\right) +\bar{s}-1\right) ,
\label{sysan}
\end{eqnarray}%
giving%
\begin{equation}
\dot{a}_{j}^{i}=a_{j}^{i-1}s_{j}^{i-1}-a_{j}^{i}\left(
\sum_{z=0}^{m}a_{j}^{z}f_{j}^{z}\left( 1-\frac{n}{K}\right)
+\sum_{z=0}^{m}a_{j}^{z}s_{j}^{z}\right) ,
\end{equation}%
\begin{equation}
\dot{p}_{j}=p_{j}\left( \left(
\sum_{i=0}^{m}a_{j}^{i}f_{j}^{i}-\sum_{z=1}^{w}p_{z}%
\sum_{i=0}^{m}a_{z}^{i}f_{z}^{i}\right) \left( 1-\frac{n}{K}\right) +\left(
\sum_{i=0}^{m}a_{j}^{i}s_{j}^{i}-\sum_{z=1}^{w}p_{z}%
\sum_{i=0}^{m}a_{z}^{i}s_{z}^{i}\right) \right) ,
\end{equation}%
\begin{equation}
\dot{n}=n\left( \sum_{j=1}^{w}p_{j}\sum_{i=0}^{m}a_{j}^{i}f_{j}^{i}\left( 1-%
\frac{n}{K}\right)
+\sum_{j=1}^{w}p_{j}\sum_{i=0}^{m}a_{j}^{i}s_{j}^{i}-1\right) .
\label{sysan+}
\end{equation}

For system $S_{b}$ we have (see Appendix D for a detailed derivation):%
\begin{eqnarray}
\dot{p}_{j}^{0} &=&\frac{1}{a^{0}}\left(
\sum_{i=0}^{m}a^{i}p_{j}^{i}f_{j}^{i}-p_{j}^{0}\bar{f}\right) \left( 1-%
\dfrac{n}{K}\right) ,  \label{simsysb1} \\
\dot{p}_{j}^{i} &=&\frac{a^{i-1}}{a^{i}}\left(
p_{j}^{i-1}s_{j}^{i-1}-p_{j}^{i}\bar{s}^{i-1}\right) ,  \label{simsysb2} \\
\dot{a}^{i} &=&a^{i-1}\bar{s}^{i-1}-a^{i}\left( \bar{f}\left( 1-\dfrac{n}{K}%
\right) +\bar{s}\right) ,  \label{simsysb3} \\
\dot{n} &=&n\left( \bar{f}\left( 1-\dfrac{n}{K}\right) +\bar{s}-1\right) .
\label{simsysbn}
\end{eqnarray}%
The expanded form of the above system will be%
\begin{equation}
\dot{p}_{j}^{0}=\frac{1}{a^{0}}\left(
\sum_{i=0}^{m}a^{i}p_{j}^{i}f_{j}^{i}-p_{j}^{0}\sum_{i=0}^{m}a^{i}%
\sum_{z=1}^{w}p_{z}^{i}f_{z}^{i}\right) \left( 1-\dfrac{n}{K}\right) ,
\label{sysb0}
\end{equation}%
\begin{equation}
\dot{p}_{j}^{i}=\frac{a^{i-1}}{a^{i}}\left(
p_{j}^{i-1}s_{j}^{i-1}-p_{j}^{i}\sum_{z=1}^{w}p_{z}^{i-1}s_{z}^{i-1}\right) ,
\label{sysb1}
\end{equation}%
\begin{equation}
\dot{a}^{i}=a^{i-1}\sum_{j=1}^{w}p_{j}^{i-1}s_{j}^{i-1}-a^{i}%
\sum_{z=0}^{m}a^{z}\left( \sum_{j=1}^{w}p_{j}^{z}f_{j}^{z}\left( 1-\dfrac{n}{%
K}\right) +\sum_{j=1}^{w}p_{j}^{z}s_{j}^{z}\right) ,  \label{sysb2}
\end{equation}%
\begin{equation}
\dot{n}=n\left( \sum_{i=0}^{m}a^{i}\left(
\sum_{j=1}^{w}p_{j}^{i}f_{j}^{i}\left( 1-\dfrac{n}{K}\right)
+\sum_{j=1}^{w}p_{j}^{i}s_{j}^{i}\right) -1\right) .  \label{sysbn}
\end{equation}%
Note that equation (\ref{sysbn}) is equivalent to (\ref{sysan+}) and in both
cases 
\begin{equation}
1-\dfrac{n}{K}=\dfrac{1-\bar{s}}{\ \bar{f}}\Rightarrow n=K\left( 1-\dfrac{1-%
\bar{s}}{\ \bar{f}}\right) .
\end{equation}%
Recall that for simplicity we assumed that functions act as the vital rates
with interaction rate $\tau $\ hidden inside. When we insert it back it
would appear as $\tau f_{j}^{i}$\ and $s_{j}^{i}=1-\tau d_{j}^{i}$. In
contrast to the basic replicator equations (\ref{QWD}), parameter $\tau $\
cannot easily be removed from systems $S_{a}$\ and $S_{b}$\ by simple
timescale adjustment. A similar situation occurs with the background payoff
components $\Phi $\ and $\Psi $, which simply cancel out in (\ref{QWD}) but
are still present in the population size equation. This will not be the case
for systems $S_{a}$\ and $S_{b}$. However, for simplicity, in this paper we
do not deal explicitly with the background payoffs.

\subsection{Mixed PDE-ODE versions of systems $S_{a}$ and $S_{b}$}

We can derive mixed PDE-ODE versions of systems $S_{a}$ and $S_{b}$, where
the age profile is a continuous function, which are simpler and more
mathematically elegant. The advantage is that they are independent of the
timescale since the interaction rate $\tau $ will simply cancel out (see
Appendix E for derivations). Thus the previous simplifying assumption about
skipping it is obsolete in this case. Payoffs $d_{j}(t,l)$ and $f_{j}(t,l)$
are now continuous functions of the lifetime $l$ and the strategic
composition at time $t$. In addition the distinction between aggregated age
class survival and game round survival discussed below equation (\ref{r}) is
not necessary since PDE versions of both systems will be driven by game
payoffs only. Therefore for system $S_{a}$ we have 
\begin{eqnarray}
\frac{\partial a_{j}(t,l)}{\partial t}+\frac{\partial a_{j}(t,l)}{\partial l}
&=&a_{j}(t,l)\left[ -d_{j}(t,l)-(\bar{f}_{j}(t)\left( 1-\frac{n(t)}{K}%
\right) -\bar{d}_{j}(t))\right] , \\
\dot{p}_{j}(t) &=&p_{j}(t)\left( \left( \bar{f}_{j}(t)-\bar{f}(t)\right)
\left( 1-\frac{n(t)}{K}\right) -\left( \bar{d}_{j}(t)-\bar{d}(t)\right)
\right) , \\
\dot{n}(t) &=&n(t)(\bar{f}(t)\left( 1-\frac{n(t)}{K}\right) -\bar{d}(t)),
\end{eqnarray}%
with $a_{j}(t,0)=\left( 1-\frac{n(t)}{K}\right) \bar{f}_{j}(t)$, $\bar{f}%
_{j}(t)=\int_{0}^{\infty }a_{j}(t,l)f_{j}(t,l)dl$, $\bar{d}%
_{j}(t)=\int_{0}^{\infty }a_{j}(t,l)d_{j}(t,l)dl$, $\bar{f}(t)=\sum_{j}$ $%
p_{j}(t)\bar{f}_{j}(t)$ and $\bar{d}(t)=\sum_{j}p_{j}(t)\bar{d}_{j}(t)$.

For system $S_{b}$ we have 
\begin{eqnarray}
\frac{\partial p_{j}(t,l)}{\partial t}+\frac{\partial p_{j}(t,l)}{\partial l}
&=&p_{j}(t,l)\left[ \bar{d}(t,l)-d_{j}(t,l)\right] , \\
\frac{\partial a(t,l)}{\partial t}+\frac{\partial a(t,l)}{\partial l}
&=&a(t,l)\left[ -\bar{d}(t,l)-(\bar{f}(t)\left( 1-\frac{n(t)}{K}\right) -%
\bar{d}(t))\right] , \\
\dot{n}(t) &=&n(t)(\bar{f}(t)\left( 1-\frac{n(t)}{K}\right) -\bar{d}(t)),
\end{eqnarray}%
with $a(t,0)=\left( 1-\frac{n(t)}{K}\right) \bar{f}(t)$, $p_{j}(t,0)=\left(
1-\frac{n}{K}\right) \int_{0}^{\infty }p_{j}(t,l)f_{j}(t,l)dl$, $\bar{d}%
(t,l)=\sum_{j}p_{j}(t)\bar{d}_{j}(t,l)$, $\bar{f}(t)=\int_{0}^{\infty }a(t,l)%
\bar{f}(t,l)dl$ and $\bar{d}(t)=\int_{0}^{\infty }a(t,l)\bar{d}(t,l)dl$,
where $\bar{f}(t,l)=\sum_{j}$ $p_{j}(t,l)\bar{f}_{j}(t,l)$ and $\bar{d}%
(t,l)=\sum_{j}p_{j}(t,l)\bar{d}_{j}(t,l)$ 

\subsection{ A sex ratio example}

Now we will show how the methods presented in the previous sections can be
used to extend the simpler age independent model to the age dependent case
and how they can be used to model a sexually reproducing population. We will
show this methodology by example of the synthetic sex ratio model
(Argasinski 2012, Argasinski 2013, Argasinski 2017) combining simple
explicit genetics (similar to the more advanced approaches as in Karlin and
Lessard 1986) with rigorous strategic analysis. We will use the formulation
of the model focused on selection of genes encoding sex ratio strategies
(Argasinski 2013). 
Below we outline the basic details of this model. The introduction of the
life cycle perspective to theoretical studies on the sex ratio is important,
since data show the huge impact age specific mortalities can have on the
dynamics of age specific sex ratios (for example see Orzack et al. 2015 for
data showing the changes of the human sex ratio from conception to death).

We have a population consisting of $x$ females and $y$ males. All of them
are carriers of a single gene encoding one from a finite number $w$ of
competing sex ratio strategies which are expressed by females (strategy $%
P_{j}\in \lbrack 0,1]$ is carried by $x_{j}$ females and $y_{j}$ males and
describes the fraction of male newborns in the brood of a female). Then the
population state can be expressed by the population's sex ratio $P=y/(x+y)$,
primary sex ratio (average strategy of females) $\bar{P}_{pr}=\sum_{j=1}^{w}%
\dfrac{x_{j}}{x}P_{j}$ and vectors $G$ and $M$ where:

$G_{j}=\dfrac{x_{j}+y_{j}}{\sum_{z=1}^{w}\left( x_{z}+y_{z}\right) }$ \ \ \
\ \ \ \ \ the gene frequencies,

$M_{j}=\dfrac{y_{j}}{x_{j}+y_{j}}$ \ \ \ the sex ratios in the carrier
subpopulations.

Then $P=\sum_{j=1}^{w}G_{j}M_{j}$ and $\bar{P}_{pr}=\sum_{j=1}^{w}\dfrac{%
x_{j}}{x}P_{j}=\sum_{j=1}^{w}\dfrac{G_{j}(1-M_{j})}{1-P}P_{j}$. \ Strategy
genes are inherited from mother or father with probability 0.5. The sex
specific payoff functions describe the impact of direct reproductive success
(offspring of the focal female or offspring of partners of the focal male)
and the per capita normlized contribution of the same strategy carriers\ of
the opposite sex. \ Therefore, the payoffs of male and female carriers and
the average gene carrier are: 
\begin{eqnarray}
f_{m}(P_{j},G,M) &=&\dfrac{k}{2}\left( \dfrac{x}{y}\bar{P}_{pr}+\dfrac{x_{j}%
}{y_{j}}P_{j}\right)  \label{eq:Wm46} \\
&=&\dfrac{k}{2}\left( \dfrac{1-P}{P}\bar{P}_{pr}+\dfrac{(1-M_{j})}{M_{j}}%
P_{j}\right) , \\
f_{f}(P_{j},G,M) &=&\dfrac{k}{2}\left( \left( 1-P_{j}\right) +\dfrac{y_{j}}{%
x_{j}}\left( 1-\bar{P}_{pr}\right) \dfrac{x}{y}\right)  \label{Wf} \\
&=&\dfrac{k}{2}\left( \left( 1-P_{j}\right) +\dfrac{M_{j}}{(1-M_{j})}\left(
1-\bar{P}_{pr}\right) \dfrac{1-P}{P}\right) , \\
f_{g}(P_{j},G,M) &=&M_{j}f_{m}(P_{j},G,M)+(1-M_{j})f_{f}(P_{j},G,M) \\
&=&\dfrac{k}{2}\left[ M_{j}\dfrac{1-P}{P}+(1-M_{j})\right]
\end{eqnarray}%
where $k$ is the number of offspring per female. The average payoffs are: 
\begin{eqnarray}
\bar{f}_{m}(G,M) &=&k\dfrac{1-P}{P}\bar{P}_{pr}, \\
\bar{f}(G,M) &=&k\left( 1-P\right) .  \label{eq:W52}
\end{eqnarray}%
We can obtain the system describing the dynamics of gene frequencies and the
sex ratios in the carrier subpopulations: 
\begin{eqnarray}
\dot{G}_{j} &=&G_{j}\left( f_{g}(P_{j},G,M)-\bar{f}(G,M)\right) ,  \label{G}
\\
\dot{M}_{j} &=&M_{j}(f_{m}(P_{j},G,M)-f_{g}(P_{j},G,M)),  \label{M}
\end{eqnarray}%
leading to the following system of equations 
\begin{eqnarray}
\dot{G}_{j} &=&G_{j}\left( \dfrac{1}{2}-P\right) \left( \dfrac{M_{j}}{P}%
-1\right) ,  \label{eqG} \\
\dot{M}_{j} &=&\dfrac{k}{2}\left( M_{j}\left( \dfrac{1-P}{P}\right) \left( 
\bar{P}_{pr}-M_{j}\right) +\left( 1-M_{j}\right) \left( P_{j}-M_{j}\right)
\right) .  \label{eqM}
\end{eqnarray}

The above system can be regarded as an example of multi-level selection
since the fate of a gene is determined by the actual composition of the
carrier subpopulation described by the carriers' sex ratio $M_{j}$ and the
threshold between growth and decline is the adult sex ratio $%
P=\sum_{j=1}^{w}G_{j}M_{j}$. The parameters $M_{j}$ are determined by the
Tug of War dynamics (\ref{eqM}) describing the impact of female carriers
producing newborns according to the carried strategy $P_{j}$ and randomly
drawn female partners of male carriers producing newborns according to the
average strategy of females $\bar{P}_{pr}$.

\subsection{The extension of the sex ratio model to the age structured case}

We will extend this system in the following way.

FIGURE 3 HERE

System $S_{a}$ will be applied to extend the gene pool dynamics to the
system with explicit age structure for each subpopulation of carriers
(described by $a_{j}^{i}$ for the $j$-th gene)\ of the particular gene. This
means that each equation (\ref{eqG}) will be transformed to the form (\ref%
{sysa2}) and completed by the respective subsystem (\ref{sysa1}) describing
the age structure of the subpopulation of carriers of the particular gene.
In addition, for the age structure of each strategy we will apply system $%
S_{b}$ to describe the dynamics of the sex ratios within each age class.
Thus for each strategy, the respective subsystem (\ref{sysa1}) will be the
core subsystem (\ref{simsysb3}) of system $S_{b}$, and it will be completed
by the respective subsystems (\ref{simsysb1},\ref{simsysb2}), describing the
dynamics of strategy carriers' sex ratios in particular age classes. This
structure will be the generalization of the $M_{j}$ equations (\ref{eqM}) in
the original model. Assume that survival, described by $s_{f}^{i}$ for
females and by $s_{m}^{i}$ \ for males, depends only on sex and age. Males
are active in the age classes from $a$ to $b$\ and females from $c$\ to $d$,
and fractions of sexually active female and male individuals carrying the $j$%
-th strategy are%
\begin{equation}
S_{j}^{f}=\sum_{z=c}^{d}a_{j}^{z}(1-M_{j}^{z}),\text{ \ \ \ \ \ \ \ }%
S_{j}^{m}=\sum_{z=a}^{b}a_{j}^{z}M_{j}^{z}\text{.}
\end{equation}%
Analogous parameters for the whole population are 
\begin{equation}
\bar{S}^{f}=\sum_{j=1}^{w}G_{j}S_{j}^{f},\hspace{0.2cm}\bar{S}%
^{m}=\sum_{j=1}^{w}G_{j}S_{j}^{m}.
\end{equation}%
We also have $P=\sum_{j=1}^{w}G_{j}\sum_{i}a_{j}^{i}M_{j}^{i}$ , and the
primary sex ratio is:%
\begin{equation}
\bar{P}_{pr}=\sum_{j=1}^{w}\dfrac{G_{j}S_{j}^{f}}{\sum_{z}G_{z}S_{z}^{f}}%
P_{j}=\dfrac{\sum_{j=1}^{w}G_{j}S_{j}^{f}P_{j}}{\bar{S}^{f}}.
\end{equation}%
Thus this is the average strategy of active females describing the
proportion of males among all newborns or zygotes. The operational sex ratio
among active carriers of strategy $j$ and the equivalent average value for
the population is 
\begin{equation}
M_{j}^{op}=\dfrac{S_{j}^{m}}{S_{j}^{m}+S_{j}^{f}},\hspace{0.2cm}P_{op}=%
\dfrac{\bar{S}^{m}}{\bar{S}^{m}+\bar{S}^{f}}.  \label{opratios}
\end{equation}%
The equations on $G$ should be updated according to the additional
assumptions on age limits of sexual activity (age classes from $a$ to $b$
for males and $c$ to $d$ for females). We should also derive the respective
forms of per capita fertility payoffs described in the new coordinates. For
derivation of the dynamics we need the following operational male fertility
payoff of active males, average per capita gene fertility payoff and the
average fertility in the whole population (the detailed derivation is in
Appendix F): \vspace{-0.1cm} 
\begin{eqnarray}
f_{m}^{op}(P_{j},a,G,M) &=&\dfrac{k}{2}\left( \dfrac{1-P_{op}}{P_{op}}\bar{P}%
_{pr}+\dfrac{1-M_{j}^{op}}{M_{j}^{op}}P_{j}\right) , \\
f_{g}(P_{j},a,G,M) &=&\dfrac{k}{2}\left( S_{j}^{f}+S_{j}^{m}\dfrac{1-P_{op}}{%
P_{op}}\right) , \\
\bar{f}(a,G,M) &=&k\bar{S}^{f}.  \label{SRfaver}
\end{eqnarray}%
Note that $\left( 1-P_{op}\right) /P_{op}$ describes the number of partners
and $\left( 1-M_{j}^{op}\right) /M_{j}^{op}$ the number of female carriers
(\textquotedblleft sisters") of the average male carrier of the focal
strategy gene. Therefore, the male operational fertility payoff $f_{m}^{op}$
describes the fertility of their partners with the average strategy and
\textquotedblleft sisters" carrying the same gene. The gene payoff $f_{g}$
describes the aggregated fertility of all female carriers and all partners
of male carriers. Thus we will obtain the following general system derived
in Appendix G: 
\begin{eqnarray}
\dot{G}_{j} &=&G_{j}\left( \left( f_{g}(P_{j},a,G,M)-\bar{f}(a,G,M)\right)
\left( 1-\dfrac{n}{K}\right) +\left( \bar{s}_{j}-\bar{s}\right) \right) ,
\label{gene1} \\
\dot{a}_{j}^{i} &=&a_{j}^{i-1}\bar{s}_{j}^{i-1}-a_{j}^{i}\left[
f_{g}(P_{j,}a,G,M)\left( 1-\dfrac{n}{K}\right) +\bar{s}_{j}\right] ,
\label{age} \\
\dot{M}_{j}^{0} &=&\dfrac{\left(
f_{m}^{op}(P_{j},a,G,M)S_{j}^{m}-M_{j}^{0}f_{g}(P_{j},a,G,M)\right) }{%
a_{j}^{0}}\left( 1-\dfrac{n}{K}\right) ,  \label{tug1} \\
\dot{M}_{j}^{i} &=&\dfrac{a_{j}^{i-1}}{a_{j}^{i}}\left(
M_{j}^{i-1}s_{m}^{i-1}-M_{j}^{i}\bar{s}_{j}^{i-1}\right) ,  \label{carr1} \\
n &=&n\left( \bar{f}\left( 1-\dfrac{n}{K}\right) +\bar{s}-1\right) ,
\label{N}
\end{eqnarray}%
where $\bar{s}_{j}^{i}=M_{j}^{i}s_{m}^{i}+\left( 1-M_{j}^{i}\right)
s_{f}^{i} $\ describes the average survival of the carrier of the $j$th
strategy determined by the actual carriers sex ratio. Then $\bar{s}%
_{j}=\sum_{i=0}^{m}a_{j}^{i}\bar{s}_{j}^{i}$\ and $\bar{s}%
=\sum_{j=1}^{w}G_{j}\bar{s}_{j}$. Thus the general equations (\ref{sysa1}-%
\ref{sysan}) have become equations (\ref{gene1}, \ref{age}, \ref{N}) through
the sequences: $(\ref{sysa2})\rightarrow (\ref{G})\rightarrow (\ref{gene1})$%
, $(\ref{sysa1})\rightarrow (\ref{simsysb3})\rightarrow (\ref{age})$, $(\ref%
{simsysb1})\rightarrow (\ref{tug1})$, $(\ref{simsysb2})\rightarrow (\ref%
{carr1})$. Figure 3 shows how the phase space of the original model was
extended to the age structured case. After substitution of the payoff
functions (see Appendix G) we obtain the system: \vspace{-0.1cm} 
\begin{eqnarray}
\dot{G}_{j} &=&G_{j}\left( k\left[ \dfrac{1}{2}\left( \dfrac{S_{j}^{f}}{\bar{%
S}^{f}}+\dfrac{S_{j}^{m}}{\bar{S}^{m}}\right) -1\right] \bar{S}^{f}\left( 1-%
\dfrac{n}{K}\right) +\left( \bar{s}_{j}-\bar{s}\right) \right) ,  \label{srg}
\\
\dot{a}_{j}^{i} &=&a_{j}^{i-1}\bar{s}_{j}^{i-1}-a_{j}^{i}\left[ \dfrac{k}{2}%
\left( S_{j}^{f}+S_{j}^{m}\dfrac{\bar{S}^{f}}{\bar{S}^{m}}\right) \left( 1-%
\dfrac{n}{K}\right) +\bar{s}_{j}\right] ,  \label{sra} \\
\dot{M}_{j}^{0} &=&\dfrac{k}{2a_{j}^{0}}\left( S_{j}^{m}\left( \bar{P}%
_{pr}-M_{j}^{0}\right) \dfrac{\bar{S}^{f}}{\bar{S}^{m}}+S_{j}^{f}\left(
P_{j}-M_{j}^{0}\right) \right) \left( 1-\dfrac{n}{K}\right) ,  \label{srm0}
\\
\dot{M}_{j}^{i} &=&\dfrac{a_{j}^{i-1}}{a_{j}^{i}}\left(
M_{j}^{i-1}s_{m}^{i-1}-M_{j}^{i}\bar{s}_{j}^{i-1}\right) ,  \label{srm} \\
n &=&n\left[ k\bar{S}^{f}\left( 1-\dfrac{n}{K}\right) +\bar{s}-1\right] ,
\label{srn}
\end{eqnarray}%
where\ average survival probabilities are\ \ 
\begin{equation}
\bar{s}_{j}^{i}=M_{j}^{i}s_{m}^{i}+(1-M_{j}^{i})s_{f}^{i},\hspace{0.2cm}\bar{%
s}_{j}=\sum_{i=1}^{m}a^{i}\bar{s}_{j}^{i}.
\end{equation}%
\begin{equation*}
S_{j}^{f}=\sum_{z=c}^{d}a_{j}^{z}(1-M_{j}^{z}),\text{ }\bar{S}%
^{f}=\sum_{j=1}^{w}G_{j}S_{j}^{f},\text{\ }S_{j}^{m}=%
\sum_{z=a}^{b}a_{j}^{z}M_{j}^{z},\text{ }\bar{S}^{m}=%
\sum_{j=1}^{w}G_{j}S_{j}^{m}
\end{equation*}%
are the fractions of sexually active females and males among the $P_{j}$
gene carriers, and the respective averages. Thus the selection mechanism is
seriously altered by the age structure. The above system shows that
differences in mortality between sexes and different ages of sexual activity
can significantly affect the selection of individual strategies. Equations (%
\ref{srm0}) contain the terms $S_{j}^{m}$ and $S_{j}^{f}$ describing the
fractions of sexually active individuals and are the equivalent of the Tug
of War dynamics (\ref{eqM}). The dynamics of the age structure of each
strategy is attracted by%
\begin{equation}
\hat{a}_{j}^{i}=\hat{a}_{j}^{i-1}\frac{M_{j}^{i-1}s_{m}^{i-1}+\left(
1-M_{j}^{i-1}\right) s_{f}^{i-1}}{\dfrac{k}{2}\left( S_{j}^{f}+S_{j}^{m}%
\dfrac{\bar{S}^{f}}{\bar{S}^{m}}\right) \left( 1-\dfrac{n}{K}\right) +\bar{s}%
_{j}}.
\end{equation}%
Sex ratios \textbf{among the }$j$\textbf{-th strategy carriers of particular
ages} converge to%
\begin{equation}
\hat{M}_{j}^{0}=\frac{\bar{P}_{pr}\dfrac{S_{j}^{m}}{\bar{S}^{m}}\bar{S}%
^{f}+S_{j}^{f}P_{j}}{\dfrac{S_{j}^{m}}{\bar{S}^{m}}\bar{S}^{f}+S_{j}^{f}},%
\hspace{0.2cm}\hat{M}_{j}^{i}=\frac{M_{j}^{i-1}s_{m}^{i-1}}{\bar{s}_{j}^{i-1}%
}\hspace{0.5cm}i>0.
\end{equation}

Note that when we assume that there are no differences in survival
probabilities between sexes ($s_{f}^{i}=s_{m}^{i}$) then the system (\ref%
{srg}-\ref{srn}) reduces to the simplified version. Equations (\ref{sra})
will be independent of parameters $M_{j}^{i}$ and the equations (\ref{srm})
will converge to a constant value over the whole life cycle ($%
M_{j}^{i}=M_{j}^{0}$ for all $i$). Therefore all strategies will have the
same age structure. In effect the bracketed term $\left( \bar{s}_{j}-\bar{s}%
\right) $, describing the excess of the mortality payoff from average
mortality, will vanish in equations (\ref{srg}) and selection of the genes
will be driven by the excess fertility payoff bracket $\left(
f_{g}(P_{j},a,G,M)-\bar{f}(a,G,M)\right) $ describing the Fisherian
mechanism driven by the difference in reproductive value between the sexes; 
\begin{equation}
\left( f_{g}(P_{j},a,G,M)-\bar{f}(a,G,M)\right) =k\left[ \dfrac{1}{2}\left( 
\dfrac{S_{j}^{f}}{\bar{S}^{f}}+\dfrac{S_{j}^{m}}{\bar{S}^{m}}\right) -1%
\right] \bar{S}^{f},  \label{Fisher}
\end{equation}%
which is equivalent to the Shaw-Mohler formula (Shaw and Mohler 1953). If we
assume that both sexes are mature in the same age classes then $%
S_{j}^{f}=1-S_{j}^{m}$, we have that operational sex ratios (\ref{opratios})
are $M_{j}^{op}=S_{j}^{m}\hspace{0.2cm}$and $P_{op}=\bar{S}^{m}$ ($\bar{S}%
^{f}=1-P_{op}$). Therefore for the operational sex ratio $P_{op}=0.5$ the
above formula equals zero for all strategies. When we additionally assume
that the sex specific survivals for different ages are the same, the system
replicates the results of the original model . In the general case if $%
S_{j}^{f}=\bar{S}^{f}$ and $S_{j}^{m}=\bar{S}^{m}$ then obviously the
operational sex ratios (\ref{opratios}) are equal. In other cases, the
strategies with the greater fraction of the sex which is in the minority
among active individuals (according to the operational sex ratio $P_{op}$)
will have a greater value of (\ref{Fisher}). Since all individuals of the
same sex suffer the same mortality, \textbf{the} values of parameters $%
S_{j}^{f}$ and $S_{j}^{m}$ are determined by the allocation of sexes at
birth, determined by their encoded strategy. Due to the constant brood size $%
k$, 
an increase of female newborns leads to a decrease of male newborns and vice
versa. Therefore, this allocation will determine operational sex ratios and
selection should act accordingly to differences in operational sex ratios,
similarly to (\ref{eqG}).

We can see this in Fig. 4 depicting a numerical simulation for the case of
three competing strategies $P_{1}=0.05$, $P_{2}=0.55$, $P_{3}=0.95$ with $25$
age classes plus infant age class 0. For simplicity we assumed that age
class survivals will be the same with only one change at some arbitrary age,
different for males and females. For females we have survival $0.95$ until
age $10$ and $0.80$ in subsequent ages. For males we have $0.88$ until age
15 and then $0.72$ subsequently. By definition survival in the last age
classes is zero. Females are fertile from age $c=8$ until age $d=15$ while
males are active from age $a=8$ until age $b=20$. The initial population
size was $n(0)=40$ with a carrying capacity $K=10000$. Initial conditions
are $G_{1}(0)=0.9$, $G_{2}(0)=G_{3}(0)=0.05$, $M_{1}^{0}=0.7$ and $%
M_{2}^{0}=M_{3}^{0}=0.1$. We start from a very young population where adult
age classes have frequencies $0.001$ leading to a $0.025$ proportion of
non-infant individuals and sex ratios are $M_{1}^{i}=0.9$ and $%
M_{2}^{i}=M_{3}^{i}=0.8$. These exaggerated conditions show the initial
dynamics of the growing cohort leading to the interesting patterns depicted
in Fig. 5 depicting the age structure and Fig 6 showing the dynamics of age
specific sex ratios.

Fig. 7 shows the delayed convergence to the respective Euler-Lotka
manifolds. At the global equilibrium excess fertility payoffs (\ref{Fisher})
(the difference between the payoff and the mean) are not equal to zero
because they must balance the nonzero values of the excess survival payoffs
for growth rates to be equal (Fig.8). Therefore, the classical Fisherian
equilibrium focused only on fertility payoffs is not reached here. A
question arises about the interplay between fertility and survival and how
it leads to the primary sex ratio of 0.5. In addition the operational sex
ratio is far from $0.5$. Therefore in this case the Fisherian mechanism is
not enough to explain the origins of the primary sex ratio of 0.5. Fig. 4
shows that the mechanism driven by the operational sex ratios still works
but all values are rescaled, and we also have different mortalities for
different strategies. The interplay between the Fisherian mechanism, driven
by fertility and differences in reproductive value between the sexes, and
age structure, driven by survival differences between the sexes, needs an
explanation which will be the subject of future work.

\vspace{0.3cm}

FIGURES 4-8 HERE

\section{Discussion}

In this work we presented a new modelling framework combining evolutionary
dynamics with demographic structure. This approach can be a useful tool in
the development of the synthesis between evolutionary game theory and life
history theory. We started with the derivation of the ODE discretized
approximation of the McKendrick von Foerster model of age structured
populations and its critical manifold equivalent to the Euler-Lotka
equation. This was extended to the explicit case of multiple competing
strategies and transformed into two types of age structured replicator
dynamics. The first focused on the selection of strategies when each
strategy is described by a subsystem describing the dynamics of the age
structure. The second focused on the age structure of the whole population
and a subsystem of the strategies within each age class. These led to huge
ODE systems which are equivalent to systems of Bernadelli-Lewis-Leslie
matrices. Another complication is that for the discretized age structure we
need age class survival functions which will describe the aggregated
outcomes of all interactions (game rounds) that have happened during a
single time unit. This survival function is distinct from game round
survival which can be used for the derivation of the fertility-survival
trade-off functions used in situations when only survivors of the
interaction can reproduce (Argasinski and Broom 2012, 2018a, 2018b).\ In
addition we have outlined the PDE versions of the obtained systems to
indicate a future direction of research. 

Both approaches are combined in the illustrative example of a sex ratio
model. This is an extension of the dynamic sex ratio model (Argasinski 2012,
2013, 2017). It shows that when we assume different mortalities for both
sexes, the classical Fisherian explanation based on the differences of
reproductive values of offspring is not enough to explain convergence to the
primary sex ratio of 0.5. The excess fertility payoff does not converge to 0
which would be equivalent to an equal reproductive values for both sexes,
but its non-zero value is equal to the value of the excess survival payoff.
The question of how this mechanism works in detail should be explained in
future research. The new model provides a theoretical framework that can be
used to explain the mechanisms shaping the patterns observable in data
collected on age specific sex ratios from conception to death, as in Orzack
et al. (2015) and Orzack (2016).

The obtained results clearly show that a life cycle perspective plays a
crucial role in evolutionary processes. In the classical approaches to
evolutionary game theory individuals cannot change their properties during
their lifetime. Thus their life history is a memoryless process, and
survival of a single interaction does not change the state of the
individual. This is caused by the fact that the classical approaches to
evolutionary games are focused on the strategies interpreted as patterns of
behaviour, not on the individual itself. The exception to this rule is the
state based approach (Houston and McNamara 1999, Argasinski and Rudnicki
2020). The explicit description of the life cycle and the different payoffs
at different ages leads to a more complicated game theoretic structure. In
particular a mixed PDE-ODE approach will lead to more complex payoff
functions based on continuous distributions of ages for different
strategies. This will need more sophisticated methods, such as models with
function valued traits (Oechssler and Riedel 2001, Dieckmann et al 2006, van
Veelen and Spreij 2009), state based games (Houston and McNamara 1991, 1999,
argasinski and Rudnicki 2020) or "large games" with a distinction between
strategy sets and population states (Wieczorek and Wiszniewska 1998,
Wieczorek 2004; 2005), as opposed to basic two person matrix games. The
modelling framework proposed in this paper can also be a useful tool in the
research on animal personalities 
The combination of game theoretic analysis with an explicit age structure
will allow us to analyze the relationships between behavioural strategies
(such as aggression or cowardice) and life history traits (such as
allocation of energy into growth or reproduction). This is important because
life history trade-offs are shaped by external mortality which is the
outcome of interactions with the environment. On the other hand, the
demographic outcomes of interactions such as mortality are affected by
phenotypic traits such as growth shaped by life history strategies (Wolf and
Weissing 2010, Wolf and McNamara 2012). This constitutes
life-history-behavioural feedback.

\section*{Appendix A: Derivation of the frequency equations}

\textbf{1) Discretization of the of McKendrick von Foerster model (\ref%
{McKen}):} We need to divide the continuous time into separate discrete unit
compartments describing age classes consisting of individuals of ages from
the interval $(l,l+1]$. Recall that game rounds occur at intensity $\tau $.
Note that the exponential dynamics emerges from an aggregation of survival
from some independent interactions, when the focal individual survives (or
not) several events and the aggregated survival is the product of survival
probabilities of those events. However we can imagine a timescale where
interactions are sufficiently rare, so that only a small fraction of
individuals play a single round of the game. Then the dynamics is
technically linear and survival is described by the first term of the Taylor
series of the exponential function. Thus assume that time interval $dt=dl=1$
is small enough that a small fraction $\tau d(t,l)dt$ of individuals will
die due to aggregated outcomes of independent game rounds. The remaining $1-$%
\ $\tau d(t,l)dt$ survivors will be moved to the next age compartment. Then
from equation (\ref{McTaylor}) we have that $n(t+1,l+1)=\left( 1-\tau
d(t,l)\right) n(t,l)=s(t,l)n(t,l)$ which describes the move from point $t,l$
to point $t+1,l+1$. Assume that during a unit interval all surviving
individuals from age $l$ will be moved to age $l+1$ while all individuals
from $l+1$ will be moved from this age to the next age step or die.
Therefore during a single time unit we have linear movement occurring at
incoming rate $s(t,l-1)n(t,l-1)$ and removal rate -$n(t,l)$, since all
individuals will be removed during a single time unit. Therefore this linear
process can be well approximated by the first order Taylor expansion for $%
\Delta t=1$ where the bracketed term can be interpreted as the first
derivative; 
\begin{equation}
n(t+1,l)=n(t,l)+[s(t,l-1)n(t,l-1)-n(t,l)],  \label{TaylorLeslie}
\end{equation}%
therefore the bracketed term constitutes derivative $dn/dt$, leading to 
\begin{equation*}
\frac{dn(t,l)}{dt}=s(t,l-1)n(t,l-1)-n(t,l).
\end{equation*}%
Note that (\ref{TaylorLeslie}) can be presented in the form $%
n(t+1,l)=s(t,l-1)n(t,l-1)$ which leads to the Leslie matrix (\ref%
{LeslieMatrix}). When we change notation to the numbered age classes
describing age increments and assume that aggregated survival rate $s^{i}(t)$
and fertility rate $\tau f^{i}\left( t\right) $ can change in time, we
obtain the system 
\begin{equation}
\dot{n}^{0}(t)=\sum_{i=0}^{m}n^{i}(t)\tau f^{i}(t)-n^{0}(t),
\label{contfert}
\end{equation}%
\begin{equation}
\dot{n}^{i}(t)=s^{i-1}(t)n^{i-1}(t)-n^{i}(t)\hspace{0.5cm}i=1,\ldots ,m.
\label{cont}
\end{equation}%
It is reasonable to assume that $s^{0}=1$, (other values are equivalent to $%
s^{0}=1$ with rescaled fertilities $f^{i}$) and $s^{m}=0$.

However, to be compatible with the replicator dynamics and game theoretic
machinery, the dynamics should be expressed in terms of phenotype
frequencies. Let us change the coordinates to the frequencies $a^{i}=n^{i}/n$
(with $n=\sum_{i=0}^{m}n^{i}$) describing the age structure. The system (\ref%
{contfert},\ref{cont}) can be presented in the form of the Malthusian
equations: 
\begin{eqnarray}
\dot{n}^{0}(t) &=&\sum_{i=0}^{m}n^{i}(t)\tau
f^{i}(t)-n^{0}(t)=n^{0}(t)\left( \sum_{i=0}^{m}\dfrac{n^{i}(t)\tau f^{i}(t)}{%
n^{0}(t)}-1\right)  \notag \\
&=&n^{0}(t)\left( \sum_{i=0}^{m}\dfrac{a^{i}(t)\tau f^{i}(t)}{a^{0}(t)}%
-1\right) ,  \label{n0(t)} \\
\dot{n}^{i}(t) &=&s^{i-1}(t)n^{i-1}(t)-n^{i}(t)=n^{i}(t)\left( \dfrac{%
n^{i-1}(t)s^{i-1}(t)}{n^{i}(t)}-1\right)  \notag \\
&=&n^{i}(t)\left( \dfrac{a^{i-1}(t)}{a^{i}(t)}s^{i-1}(t)-1\right) \hspace{%
0.5cm}i=1,\ldots ,m.  \label{age structure}
\end{eqnarray}%
Therefore, this system can be presented as a system of frequency dependent
replicator equations $\dot{a}^{i}=a^{i}(r^{i}-\bar{r})$ and a single
equation on the scaling parameter $\dot{n}=n\bar{r}$. Since $%
\sum_{i=0}^{m}a^{i}=1$ and $s^{m}=0$ we have the average Malthusian growth
rate as 
\begin{equation}
\bar{r}=a^{0}\left( \sum_{i=0}^{m}\dfrac{a^{i}\tau f^{i}}{a^{0}}-1\right)
+\sum_{i=1}^{m}a^{i}\left( \dfrac{a^{i-1}s^{i-1}}{a^{i}}-1\right)
=\sum_{i=0}^{m}a^{i}\left( \tau f^{i}+s^{i}\right) -1.
\end{equation}
Then we can formulate a system of frequency dependent replicator equations
by transforming equation (\ref{age structure}) for $i=1,\ldots ,m$ (the
equation for $0$ is redundant and can be removed and $a^{0}=1-\sum%
\limits_{i=1}^{m}a^{i}$): 
\begin{eqnarray}
\dot{a}^{i} &=&a^{i}\left( \dfrac{a^{i-1}}{a^{i}}s^{i-1}-1-\bar{r}\right)
=a^{i-1}s^{i-1}-a^{i}\left( \sum_{k=0}^{m}a^{k}\left( \tau
f^{k}+s^{k}\right) \right)  \label{age frequencies} \\
\dot{n} &=&n\bar{r}=n\left( \sum_{i=0}^{m}a^{i}\left( \tau
f^{i}+s^{i}\right) -1\right) .
\end{eqnarray}%
To add density dependence we should multiply the fertility rate by the
logistic suppression coefficient $\left( 1-\dfrac{n}{K}\right) $.

2) \textbf{Frequency and density dependence and the choice of time unit
determining the discretization step: }When the above system describes the
dynamics of the age structure of the subpopulation of carriers of some
strategy competing with other strategies (indexed by subscripts) then the
parameters $s_{j}^{i}(t)=1-\tau d_{j}^{i}(p(t),a(t))$ and $\tau
f_{j}^{i}\left( p(t),a(t)\right) $ (thus $r_{j}=\tau \tilde{r}_{j}=\tau
\sum_{i=0}^{m}a_{j}^{i}\left( f_{j}^{i}\left( 1-\dfrac{n}{K}\right)
-d_{j}^{i}\right) $)\ are game payoffs depending on strategy frequencies $%
p_{j}=n_{j}/\sum_{k}n_{k}$ (and their age distributions, but now we limit
our reasoning to strategy frequencies $p$). We want to choose the longest
possible time step to reduce the number of equations. However the
frequencies will change in time, so the discretization step cannot be too
big, since the aggregated payoffs during unit interval will depend on the
changes of $\tau d_{j}^{i}(p(t))$ and $\tau f_{j}^{i}(p(t))$ during that
time interval. The time unit should be short enough that these vital rates
will not change significantly and the number of individuals will change
nearly linearly within each age class. Thus we should analyze how much the
strategy frequencies $p_{j}$ can change during unit time and how this
affects the vital rates. For small $\Delta t=1$ we have a change of%
\begin{equation*}
\Delta n_{j}=n_{j}(t)\tau \tilde{r}_{j}(t)\Delta t=\sum_{i}n_{j}^{i}(t)\tau
\left( f_{j}^{i}(p(t))\left( 1-\dfrac{n}{K}\right) -d_{j}^{i}(p(t))\right)
\Delta t\ 
\end{equation*}%
(positive or negative) for each $j$, leading to the change $\Delta
n=\sum_{j}\Delta n_{j}=n(t)\tau \tilde{r}(t)\Delta t$ for the population
size. Then 
\begin{eqnarray*}
p_{j}(t+1) &=&\frac{n_{j}(t)}{n(t)+\Delta n}+\frac{\Delta n_{j}}{n(t)+\Delta
n}=\frac{n_{j}(t)}{n(t)}\frac{n(t)}{n(t)+\Delta n}+\frac{\Delta n_{j}}{%
n(t)+\Delta n} \\
&=&\frac{n_{j}(t)}{n(t)}\left[ 1-\frac{\Delta n}{n(t)+\Delta n}\right] +%
\frac{\Delta n_{j}}{n(t)+\Delta n}=p_{j}(t)+\frac{\Delta
n_{j}-p_{j}(t)\Delta n}{n(t)+\Delta n}.
\end{eqnarray*}%
Therefore $p_{j}(t+1)=p_{j}(t)+\Delta p_{j}$ where vector $\Delta p$
consists of 
\begin{equation}
\Delta p_{j}(\tau )=\dfrac{\Delta n_{j}-p_{j}(t)\Delta n}{n(t)+\Delta n}=%
\dfrac{\tau }{1+\tau \tilde{r}(t))}p_{j}(t)\left( \tilde{r}_{j}(t)-\tilde{r}%
(t)\right) .  \label{deltap}
\end{equation}%
Thus the pace of increment is nearly linear or slower since $\tau /(1+\tau 
\tilde{r}(t))<\tau $. We can set the timescale by adjusting parameter $\tau $
in the formulae $\Delta f_{j}^{i}=f_{j}^{i}(p(t)+\Delta p(\tau
))-f_{j}^{i}(p(t))$ and $\Delta d_{j}^{i}=d_{j}^{i}(p(t)+\Delta p(\tau
))-d_{j}^{i}(p(t))$ to make their values small enough that $\tau \Delta
f_{j}^{i}$ and $\tau \Delta d_{j}^{i}$ are negligible. For the density
factor we have that $\left( 1-n(t+1)/K\right) =\left( 1-(n(t)+\Delta
n)/K\right) $ and the change of fertility rate is $-\tau
f_{j}^{i}(p(t))\Delta n/K$, thus it depends on $\tau /K$ and is negligible
for even big time steps.

\section*{ Appendix B: The stationary age distribution and the Euler-Lotka
equation in the continuous case}

From (\ref{age frequencies}) (recall that $\tau $ is hidden in the
fertilities $f^{i}$),\ for the non-infant age classes ($i>0$) the stationary
points for the age structure of this system are: 
\begin{equation}
\dfrac{\hat{a}^{i-1}s^{i-1}}{\hat{a}^{i}}=\sum_{i=0}^{m}\hat{a}^{i}\left(
f^{i}+s^{i}\right) \hspace{0.5cm}i=1,\ldots ,m,
\end{equation}%
therefore 
\begin{equation}
\hat{a}^{i}=\dfrac{\hat{a}^{i-1}s^{i-1}}{\sum_{i=0}^{m}\hat{a}^{i}\left(
f^{i}+s^{i}\right) }=\dfrac{\hat{a}^{i-1}s^{i-1}}{\bar{r}(\hat{a})+1}%
\Rightarrow
\end{equation}%
%
%
%
%
%
%
%
%
%
%
%
\begin{equation}
\hat{a}^{i}=\dfrac{\hat{a}^{0}\prod\limits_{z=0}^{i-1}s^{z}}{\left( \bar{r}(%
\hat{a})+1\right) ^{i}}.  \label{eq:stablefreq}
\end{equation}%
Then $\sum_{i=0}^{m}\hat{a}^{i}=1$ implies that 
\begin{equation}
\hat{a}^{0}\left( \sum_{i=0}^{m}\dfrac{\prod\limits_{z=0}^{i-1}s^{z}}{\left( 
\bar{r}(\hat{a})+1\right) ^{i}}\right) =1\Rightarrow \hat{a}^{0}=\left(
\sum_{i=0}^{m}\dfrac{\prod\limits_{z=0}^{i-1}s^{z}}{\left( \bar{r}(\hat{a}%
)+1\right) ^{i}}\right) ^{-1}.
\end{equation}%
(note the similarity to the Euler-Lotka equation). The stable age structure
is a unique vector of frequencies among age classes, conditional on the
average Malthusian growth rate of the population. Now let us prove the
equivalence with the Euler-Lotka equation. After substitution of the stable
age frequencies from equation (\ref{eq:stablefreq}) into equation (\ref%
{n0(t)}) 
we obtain: 
\begin{equation}
\dot{n}^{0}=n^{0}\left( \sum_{i=0}^{m}\dfrac{\hat{a}^{i}f^{i}}{\hat{a}^{0}}%
-1\right) =n^{0}\left( \sum_{i=0}^{m}\dfrac{f^{i}\prod%
\limits_{z=0}^{i-1}s^{z}}{\left( \bar{r}(\hat{a})+1\right) ^{i}}-1\right) .
\end{equation}%
Frequency equilibrium implies that per capita growth rates in all age
classes are equal to the average growth rate $\bar{r}(\hat{a})$. 
Thus equality will also be satisfied for the growth rate of the $0$ age
class, leading to 
\begin{equation}
\sum_{i=0}^{m}\dfrac{f^{i}\prod\limits_{z=0}^{i-1}s^{z}}{\left( \bar{r}(\hat{%
a})+1\right) ^{i}}-1=\bar{r}(\hat{a})\Rightarrow \sum_{i=0}^{m}\dfrac{%
f^{i}\prod\limits_{z=0}^{i-1}s^{z}}{\left( \bar{r}(\hat{a})+1\right) ^{i+1}}%
=1,
\end{equation}%
%
%
%
%
%
%
%
%
%
%
%
%
%
%
%
%
%
%
%
%
%
%
%
%
%
%
%
%
%
%
%
%
%
%
%
%
%
%
%
%
%
%
%
%
%
which is the Euler-Lotka equation.

\section*{ Appendix C: Derivation of system $S_{a}$}

We start from the Malthusian system describing the dynamics of age classes
in the subpopulation of carriers of the $j$-th strategy: 
\begin{eqnarray}
\dot{n}_{j}^{0} &=&\sum_{i=0}^{m}n_{j}^{i}f_{j}^{i}-n_{j}^{0}, \\
\dot{n}_{j}^{i} &=&s_{j}^{i-1}n_{j}^{i-1}-n_{j}^{i}.
\end{eqnarray}%
According to (\ref{agefreq}) the above system can be transformed into the
frequency replicator dynamics of age classes: 
\begin{equation}
\dot{a}_{j}^{i}=a_{j}^{i-1}s_{j}^{i-1}-a_{j}^{i}\left(
\sum_{k=0}^{m}a_{j}^{k}\left( f_{j}^{k}+s_{j}^{k}\right) \right) .
\end{equation}%
The Malthusian equation describing the growth of $j$-strategists is 
\begin{equation}
\dot{n}_{j}=n_{j}\bar{r}_{j}(a_{j})=n_{j}\left(
\sum_{i=0}^{m}a_{j}^{i}\left( f_{j}^{i}+s_{j}^{i}\right) -1\right) .
\end{equation}%
Then the replicator dynamics for the changes of strategy frequencies are 
\begin{equation}
\dot{p}_{j}=p_{j}\left( \bar{r}_{j}-\bar{r}\right) =p_{j}\left( \left( \bar{f%
}_{j}-\bar{f}\right) +(\bar{s}_{j}-\bar{s})\right) ,
\end{equation}%
where $\bar{r}=\sum_{j=1}^{w}p_{j}\bar{r}_{j}$ , $\bar{f}_{j}=%
\sum_{i=0}^{m}a_{j}^{i}f_{j}^{i}$, $\bar{f}=\sum_{j=1}^{w}p_{j}\bar{f}_{j}$, 
$\bar{s}_{j}=\sum_{i=0}^{m}a_{j}^{i}s_{j}^{i}$ and $\bar{s}%
=\sum_{j=1}^{w}p_{j}\bar{s}_{j}$. This gives 
\begin{eqnarray}
\dot{p}_{j} &=&p_{j}\left( \sum_{i=0}^{m}a_{j}^{i}\left(
f_{j}^{i}+s_{j}^{i}\right) -\sum_{z=1}^{w}p_{z}\sum_{i=0}^{m}a_{z}^{i}\left(
f_{z}^{i}+s_{z}^{i}\right) \right) = \\
&&p_{j}\left( \left(
\sum_{i=0}^{m}a_{j}^{i}f_{j}^{i}-\sum_{z=1}^{w}p_{z}%
\sum_{i=0}^{m}a_{z}^{i}f_{z}^{i}\right) +\left(
\sum_{i=0}^{m}a_{j}^{i}s_{j}^{i}-\sum_{z=1}^{w}p_{z}%
\sum_{i=0}^{m}a_{z}^{i}s_{z}^{i}\right) \right).
\end{eqnarray}%
The equation on the scaling parameter is 
\begin{equation}
\dot{n}=n\bar{r}=n\left( \sum_{j=1}^{w}p_{j}\sum_{i=0}^{m}a_{j}^{i}\left(
f_{j}^{i}+s_{j}^{i}\right) -1\right) .
\end{equation}%
Then to add neutral density dependence the fertilities $f_{j}^{i}$ should be
multiplied by the logistic suppression coefficient $\left( 1-n/K\right) $
leading to the system: 
\begin{eqnarray}
\dot{a}_{j}^{i} &=&a_{j}^{i-1}s_{j}^{i-1}-a_{j}^{i}\left( \bar{f}_{j}\left(
1-\frac{n}{K}\right) +\bar{s}_{j}\right) , \\
\dot{p}_{j} &=&p_{j}\left( \left( \bar{f}_{j}-\bar{f}\right) \left( 1-\frac{n%
}{K}\right) +\left( \bar{s}_{j}-\bar{s}\right) \right) , \\
\dot{n} &=&n\left( \bar{f}\left( 1-\frac{n}{K}\right) +\bar{s}-1\right) ,
\end{eqnarray}%
giving%
\begin{equation}
\dot{a}_{j}^{i}=a_{j}^{i-1}s_{j}^{i-1}-a_{j}^{i}\left(
\sum_{z=0}^{m}a_{j}^{z}f_{j}^{z}\left( 1-\frac{n}{K}\right)
+\sum_{z=0}^{m}a_{j}^{z}s_{j}^{z}\right),
\end{equation}%
\begin{equation}
\dot{p}_{j}=p_{j}\left( \left(
\sum_{i=0}^{m}a_{j}^{i}f_{j}^{i}-\sum_{z=1}^{w}p_{z}%
\sum_{i=0}^{m}a_{z}^{i}f_{z}^{i}\right) \left( 1-\frac{n}{K}\right) +\left(
\sum_{i=0}^{m}a_{j}^{i}s_{j}^{i}-\sum_{z=1}^{w}p_{z}%
\sum_{i=0}^{m}a_{z}^{i}s_{z}^{i}\right) \right) ,
\end{equation}%
\begin{equation}
\dot{n}=n\left( \sum_{j=1}^{w}p_{j}\sum_{i=0}^{m}a_{j}^{i}f_{j}^{i}\left( 1-%
\frac{n}{K}\right)
+\sum_{j=1}^{w}p_{j}\sum_{i=0}^{m}a_{j}^{i}s_{j}^{i}-1\right) .
\end{equation}

\section*{ Appendix D: Derivation of system $S_{b}$}

System $S_{b}$ produces more complex equations. As in Appendix C, to add
neutral density dependence the fertilities should be multiplied by the
logistic suppression coefficient $\left( 1-\dfrac{n}{K}\right) $. Again we
start from the Malthusian equations 
\begin{eqnarray}
\dot{n}_{j}^{0} &=&\sum_{i=0}^{m}n_{j}^{i}f_{j}^{i}\left( 1-\dfrac{n}{K}%
\right) -n_{j}^{0}=n_{j}^{0}\left( \sum_{i=0}^{m}\frac{n_{j}^{i}}{n_{j}^{0}}%
f_{j}^{i}\left( 1-\dfrac{n}{K}\right) -1\right) ,  \label{bmalt0} \\
\dot{n}_{j}^{i} &=&s_{j}^{i-1}n_{j}^{i-1}-n_{j}^{i}=n_{j}^{i}\left(
s_{j}^{i-1}\frac{n_{j}^{i-1}}{n_{j}^{i}}-1\right) .  \label{bmalt}
\end{eqnarray}%
Deriving the equations for the growth of age classes in the global
population in coordinates 
\begin{equation}
n^{i}=\sum_{j=1}^{w}n_{j}^{i},\hspace{0.2cm}p_{j}^{i}=\dfrac{n_{j}^{i}}{n^{i}%
},
\end{equation}%
we obtain%
\begin{equation}
\dot{n}^{0}=\sum_{j=1}^{w}\left( \sum_{i=0}^{m}n_{j}^{i}f_{j}^{i}\left( 1-%
\dfrac{n}{K}\right) -n_{j}^{0}\right)
=\sum_{j=1}^{w}\sum_{i=0}^{m}n_{j}^{i}f_{j}^{i}\left( 1-\dfrac{n}{K}\right)
-\sum_{j=1}^{w}n_{j}^{0}.
\end{equation}%
Since $n^{0}=\sum_{j=1}^{w}n_{j}^{0}$ and 
\begin{eqnarray}
\sum_{j=1}^{w}\sum_{i=0}^{m}n_{j}^{i}f_{j}^{i}\left( 1-\dfrac{n}{K}\right) 
&=&\sum_{i=0}^{m}n^{i}\sum_{j=1}^{w}p_{j}^{i}f_{j}^{i}\left( 1-\dfrac{n}{K}%
\right) = \\
n^{0}\sum_{i=0}^{m}\dfrac{n^{i}}{n^{0}}\sum_{j=1}^{w}p_{j}^{i}f_{j}^{i}%
\left( 1-\dfrac{n}{K}\right)  &=&n^{0}\sum_{i=0}^{m}\dfrac{a^{i}}{a^{0}}%
\sum_{j=1}^{w}p_{j}^{i}f_{j}^{i}\left( 1-\dfrac{n}{K}\right) ,
\end{eqnarray}%
the above equation has form: 
\begin{equation}
\dot{n}^{0}=n^{0}\left( \sum_{i=0}^{m}\dfrac{a^{i}}{a^{0}}%
\sum_{j=1}^{w}p_{j}^{i}f_{j}^{i}\left( 1-\dfrac{n}{K}\right) -1\right) .
\label{sbn0}
\end{equation}%
Analogously we have $n^{i}=\sum_{j=1}^{w}n_{j}^{i}$ and 
\begin{equation}
\dot{n}^{i}=\sum_{j=1}^{w}\left( s_{j}^{i-1}n_{j}^{i-1}-n_{j}^{i}\right)
=n^{i-1}\sum_{j=1}^{w}p_{j}^{i-1}s_{j}^{i-1}-n^{i}.  \label{eq139}
\end{equation}%
In the new coordinates the equation for the population size will be : 
\begin{eqnarray}
\dot{n} &=&\sum_{i=0}^{m}\dot{n}^{i}=\sum_{i=0}^{m}n^{i}%
\sum_{j=1}^{w}p_{j}^{i}f_{j}^{i}\left( 1-\dfrac{n}{K}\right)
-n^{0}+\sum_{i=1}^{m}\left(
n^{i-1}\sum_{j=1}^{w}p_{j}^{i-1}s_{j}^{i-1}-n^{i}\right)   \notag \\
&=&n\sum_{i=0}^{m}a^{i}\sum_{j=1}^{w}p_{j}^{i}f_{j}^{i}\left( 1-\dfrac{n}{K}%
\right) +n\sum_{i=1}^{m}a^{i-1}\sum_{j=1}^{w}p_{j}^{i-1}s_{j}^{i-1}-n,
\end{eqnarray}%
and since $s^{m}=0$ we can denote the above equation as 
\begin{equation}
\dot{n}=n\left( \bar{f}\left( 1-\dfrac{n}{K}\right) +\bar{s}-1\right)
=n\left( \sum_{i=0}^{m}a^{i}\sum_{j=1}^{w}p_{j}^{i}f_{j}^{i}\left( 1-\dfrac{n%
}{K}\right) +\sum_{i=0}^{m}a^{i}\sum_{j=1}^{w}p_{j}^{i}s_{j}^{i}-1\right) .
\label{sbn}
\end{equation}%
Therefore (\ref{sbn0},\ref{eq139},\ref{sbn}) can be presented as the
replicator dynamics (\ref{agefreq},\ref{agescale}) 
\begin{equation}
\dot{a}^{i}=a^{i-1}\sum_{j=1}^{w}p_{j}^{i-1}s_{j}^{i-1}-a^{i}%
\sum_{z=0}^{m}a^{z}\left( \sum_{j=1}^{w}p_{j}^{z}f_{j}^{z}\left( 1-\dfrac{n}{%
K}\right) +\sum_{j=1}^{w}p_{j}^{z}s_{j}^{z}\right) ,
\end{equation}%
\begin{equation}
\dot{n}=n\left( \sum_{i=0}^{m}a^{i}\left(
\sum_{j=1}^{w}p_{j}^{i}f_{j}^{i}\left( 1-\dfrac{n}{K}\right)
+\sum_{j=1}^{w}p_{j}^{i}s_{j}^{i}\right) -1\right) .
\end{equation}%
The above system for each age class should be completed by the replicator
dynamics describing the changes of strategic composition of this particular
age class. Since $a^{i}p_{j}^{i}=n_{j}^{i}/n$ we have the following form of (%
\ref{bmalt0}):%
\begin{equation}
\dot{n}_{j}^{0}=n_{j}^{0}\left( \dfrac{\sum_{i=0}^{m}a^{i}p_{j}^{i}f_{j}^{i}%
}{a^{0}p_{j}^{0}}\left( 1-\dfrac{n}{K}\right) -1\right) \Rightarrow 
\end{equation}%
\vspace{-0.7cm} 
\begin{eqnarray}
\dot{p}_{j}^{0} &=&p_{j}^{0}\left( \dfrac{%
\sum_{i=0}^{m}a^{i}p_{j}^{i}f_{j}^{i}}{a^{0}p_{j}^{0}}%
-\sum_{z=1}^{w}p_{z}^{0}\dfrac{\sum_{i=0}^{m}a^{i}p_{z}^{i}f_{z}^{i}}{%
a^{0}p_{z}^{0}}\right) \left( 1-\dfrac{n}{K}\right)  \\
&=&\frac{1}{a^{0}}\left(
\sum_{i=0}^{m}a^{i}p_{j}^{i}f_{j}^{i}-p_{j}^{0}\sum_{i=0}^{m}a^{i}%
\sum_{z=1}^{w}p_{z}^{i}f_{z}^{i}\right) \left( 1-\dfrac{n}{K}\right) ;
\end{eqnarray}%
analogously for other age classes we have that (\ref{bmalt}) can be
presented as%
\begin{equation}
\dot{n}_{j}^{i}=n_{j}^{i}\left( s_{j}^{i-1}\frac{a^{i-1}p_{j}^{i-1}}{%
a^{i}p_{j}^{i}}-1\right) ,
\end{equation}%
leading to the replicator dynamics%
\begin{eqnarray}
\dot{p}_{j}^{i} &=&p_{j}^{i}\left( \frac{a^{i-1}p_{j}^{i-1}}{a^{i}p_{j}^{i}}%
s_{j}^{i-1}-\sum_{z=1}^{w}p_{z}^{i}\frac{a^{i-1}p_{z}^{i-1}}{a^{i}p_{z}^{i}}%
s_{z}^{i-1}\right)  \\
&=&\frac{a^{i-1}}{a^{i}}\left(
p_{j}^{i-1}s_{j}^{i-1}-p_{j}^{i}\sum_{z=1}^{w}p_{z}^{i-1}s_{z}^{i-1}\right) .
\end{eqnarray}%
Now system $S_{b}$ can be completed%
\begin{equation}
\dot{p}_{j}^{0}=\frac{1}{a^{0}}\left(
\sum_{i=0}^{m}a^{i}p_{j}^{i}f_{j}^{i}-p_{j}^{0}\sum_{i=0}^{m}a^{i}%
\sum_{z=1}^{w}p_{z}^{i}f_{z}^{i}\right) \left( 1-\dfrac{n}{K}\right) ,
\end{equation}%
\begin{equation}
\dot{p}_{j}^{i}=\frac{a^{i-1}}{a^{i}}\left(
p_{j}^{i-1}s_{j}^{i-1}-p_{j}^{i}\sum_{z=1}^{w}p_{z}^{i-1}s_{z}^{i-1}\right) ,
\end{equation}%
\begin{equation}
\dot{a}^{i}=a^{i-1}\sum_{j=1}^{w}p_{j}^{i-1}s_{j}^{i-1}-a^{i}%
\sum_{z=0}^{m}a^{z}\left( \sum_{j=1}^{w}p_{j}^{z}f_{j}^{z}\left( 1-\dfrac{n}{%
K}\right) +\sum_{j=1}^{w}p_{j}^{z}s_{j}^{z}\right) ,
\end{equation}%
\begin{equation}
\dot{n}=n\left( \sum_{i=0}^{m}a^{i}\left(
\sum_{j=1}^{w}p_{j}^{i}f_{j}^{i}\left( 1-\dfrac{n}{K}\right)
+\sum_{j=1}^{w}p_{j}^{i}s_{j}^{i}\right) -1\right) .
\end{equation}%
This is the system (\ref{simsysb1},\ref{simsysb2},\ref{simsysb3},\ref%
{simsysbn}).

\section{Appendix E: Rescaling the McKendrick von Foerster model to
frequencies}

We can rescale the McKendrick von Foerster (\ref{McKen}) equation to
relative frequencies $a(t,l)=n(t,l)/n(t)$ where $n(t)=\int_{0}^{\infty
}n(t,l)dl$ is the size of the whole population. \textbf{In this case the
interaction rate }$\tau $ \textbf{should be explicitly considered.} Note
that in the game theoretic applications the vital rates $\tau d(t,l)$\ and $%
\tau f(t,l)$\ will change in time due to the changes of the strategic
population composition. Since $n(t,l)=a(t,l)n(t)$ and the dynamics of the
population size satisfies the Malthusian equation $\dfrac{\partial n(t)}{%
\partial t}=n(t)\bar{r}(t)=n(t)\tau \tilde{r}(t)$ equation (\ref{McKen}) can
be presented as%
\begin{equation*}
\frac{\partial a(t,l)}{\partial t}n(t)+\frac{\partial n(t)}{\partial t}%
a(t,l)+\frac{\partial a(t,l)}{\partial l}n(t)=-\tau d(t,l)a(t,l)n(t).
\end{equation*}%
After substitution $\dfrac{\partial n(t)}{\partial t}=n(t)\tau \tilde{r}(t)$
and division by $n(t)$ we obtain 
\begin{equation}
\frac{\partial a(t,l)}{\partial t}+\frac{\partial a(t,l)}{\partial l}=a(t,l)%
\left[ -\tau d(t,l)-\tau \tilde{r}(t)\right] ,  \label{frMcKen}
\end{equation}%
where per capita growth rate $\tau \tilde{r}=\tau \bar{f}(t)\left( 1-\frac{%
n(t)}{K}\right) -\tau \bar{d}(t)=$ \newline
$\tau \left[ \left( 1-\frac{n(t)}{K}\right) \int_{0}^{\infty
}a(t,l)f(l)dl-\int_{0}^{\infty }n(t,l)d(l)dl\right] $ and the boundary
condition will be $a(t,0)=n(t,0)/n(t)=\tau \left( 1-\frac{n(t)}{K}\right)
\int_{0}^{\infty }a(t,l)f(t,l)dl$. Note that we can remove $\tau $ from
equation (\ref{frMcKen}) by a change of timescale. We can use equation (\ref%
{frMcKen}) as the PDE equivalent of equations (\ref{sysa1}) from system $%
S_{a}$ and (\ref{simsysb3}) from $S_{b}$, and rescale equation (\ref{McKen})
for $j$th strategy to $p_{j}(t,l)=n_{j}(t,l)/n(t,l)\,\ $(where $%
n(t,l)=\sum_{j}n_{j}(t,l)$) to obtain the PDE equivalent of equations (\ref%
{simsysb1},\ref{simsysb2}). Then since $n_{j}(t,l)=p_{j}(t,l)n(t,l)$, 
\begin{equation*}
\frac{\partial p_{j}(t,l)}{\partial t}n(t,l)+\frac{\partial n(t,l)}{\partial
t}p_{j}(t,l)+\frac{\partial p_{j}(t,l)}{\partial l}n(t,l)+\frac{\partial
n(t,l)}{\partial l}p_{j}(t,l)=-\tau d_{j}(t,l)p_{j}(t,l)n(t,l),
\end{equation*}%
which leads to 
\begin{equation*}
\left[ \frac{\partial p_{j}(t,l)}{\partial t}+\frac{\partial p_{j}(t,l)}{%
\partial l}\right] n(t,l)+\left[ \frac{\partial n(t,l)}{\partial t}+\frac{%
\partial n(t,l)}{\partial l}\right] p_{j}(t,l)=-\tau
d_{j}(t,l)p_{j}(t,l)n(t,l).
\end{equation*}%
Substituting using equation (\ref{McKen}) 
describing whole population and division by $n(t,l)$ leads to: 
\begin{equation*}
\frac{\partial p_{j}(t,l)}{\partial t}+\frac{\partial p_{j}(t,l)}{\partial l}%
=-\tau p_{j}(t,l)\left[ d_{j}(t,l)-\bar{d}(t,l)\right] ,
\end{equation*}%
and the boundary condition replacing equation (\ref{simsysb1}) is $%
p_{j}(t,0)=n_{j}(t,0)/n(t,0)=\tau \left( 1-\frac{n(t)}{K}\right)
\int_{0}^{\infty }p_{j}(t,l)f_{j}(t,l)dl$. Parameter $\tau $ can easily be
removed in the resulting equivalents of systems $S_{a}$ and $S_{b}$ by the
timescale adjustment, thus the continuous framework is timescale independent
and can be driven by the game demographic payoff functions.

\section*{ Appendix F: Derivation of the payoff functions for the age
structured sex ratio model}

Our operational payoff functions of active individuals (\ref{eq:Wm46}) and (%
\ref{Wf}) can be presented in new coordinates in the following way: 
\begin{eqnarray}
f_{m}^{op}(P_{j},a,G,M) &=&\dfrac{k}{2}\left( \dfrac{x}{y}\bar{P}_{pr}+%
\dfrac{x_{j}}{y_{j}}P_{j}\right) , \\
&=&\dfrac{k}{2}\left( \dfrac{1-P_{op}}{P_{op}}\bar{P}_{pr}+\dfrac{%
1-M_{j}^{op}}{M_{j}^{op}}P_{j}\right) , \\
f_{f}^{op}(P_{j},a,G,M) &=&\dfrac{k}{2}\left( \left( 1-P_{j}\right) +\dfrac{%
y_{j}}{x_{j}}\left( 1-\bar{P}_{pr}\right) \dfrac{x}{y}\right) , \\
&=&\dfrac{k}{2}\left( \left( 1-P_{j}\right) +\dfrac{M_{j}^{op}}{1-M_{j}^{op}}%
\left( 1-\bar{P}_{pr}\right) \dfrac{1-P_{op}}{P_{op}}\right) .
\end{eqnarray}%
Note that in the age structured case $x$, $y$, $x_{j}$ and $y_{j}$ describe
the numbers of sexually active individuals of both sexes. Males are active
in age classes from $a$ to $b$ and females from $c$ to $d$. Fractions of
sexually active females and males in the new formulation can be presented in
the form: 
\begin{equation}
S_{j}^{f}=\sum_{i=c}^{d}a_{j}^{i}(1-M_{j}^{i})\text{ \ \ \ \ \ and \ \ \ \ \
\ \ }S_{j}^{m}=\sum_{i=a}^{b}a_{j}^{i}M_{j}^{i}.
\end{equation}%
Then $\bar{S}^{f}=\sum_{j=1}^{w}G_{j}S_{j}^{f}$ and$\hspace{0.2cm}\bar{S}%
^{m}=\sum_{j=1}^{w}G_{j}S_{j}^{m}$ are the respective averages, and
operational sex ratios are $M_{j}^{op}=\dfrac{S_{j}^{m}}{S_{j}^{m}+S_{j}^{f}}%
\hspace{0.2cm}$and $P_{op}=\dfrac{\bar{S}^{m}}{\bar{S}^{m}+\bar{S}^{f}}$.
Then the operational fertility payoff function of a gene carrier will be 
\begin{gather}
f_{g}^{op}(P_{j},a,G,M)=M_{j}^{op}f_{m}^{op}(P_{j},a,G,M)+\left(
1-M_{j}^{op}\right) f_{f}^{op}(P_{j},a,G,M)  \notag \\
=\dfrac{k}{2}\left( M_{j}^{op}\left( \dfrac{1-P_{op}}{P_{op}}\bar{P}_{pr}+%
\dfrac{1-M_{j}^{op}}{M_{j}^{op}}P_{j}\right) +\left( 1-M_{j}^{op}\right)
\left( \left( 1-P_{j}\right) +\dfrac{M_{j}^{op}}{1-M_{j}^{op}}\left( 1-\bar{P%
}_{pr}\right) \dfrac{1-P_{op}}{P_{op}}\right) \right)   \notag \\
=\dfrac{k}{2}\left( M_{j}^{op}\dfrac{1-P_{op}}{P_{op}}\bar{P}_{pr}+\left(
1-M_{j}^{op}\right) P_{j}+\left( 1-M_{j}^{op}\right) \left( 1-P_{j}\right)
+M_{j}^{op}\left( 1-\bar{P}_{pr}\right) \dfrac{1-P_{op}}{P_{op}}\right)  
\notag \\
=\dfrac{k}{2}\left( \left( 1-M_{j}^{op}\right) +M_{j}^{op}\dfrac{1-P_{op}}{%
P_{op}}\right) .
\end{gather}%
To obtain the per capita gene carrier fertility payoff, necessary to derive
the replicator equations, we multiply the above by the fraction of active
carriers 
\begin{equation}
\left[ S_{j}^{m}+S_{j}^{f}\right] =\left[ \sum_{i=a}^{b}a_{j}^{i}M_{j}^{i}+%
\sum_{i=c}^{d}a_{j}^{i}\left( 1-M_{j}^{i}\right) \right] 
\end{equation}%
leading to 
\begin{eqnarray}
f_{g}(P_{j},a,G,M) &=&\dfrac{k}{2}\left( \left( 1-M_{j}^{op}\right)
+M_{j}^{op}\dfrac{1-P_{op}}{P_{op}}\right) \left[ S_{j}^{m}+S_{j}^{f}\right] 
\\
&=&\dfrac{k}{2}\left( S_{j}^{f}+S_{j}^{m}\dfrac{1-P_{op}}{P_{op}}\right)  \\
&=&\dfrac{k}{2}\left( S_{j}^{f}+S_{j}^{m}\dfrac{\bar{S}^{f}}{\bar{S}^{m}}%
\right) .
\end{eqnarray}%
Similarly we derive the per capita average fertility. Then the average
payoff (\ref{eq:W52}) becomes the operational average fertility of the
active individuals 
\begin{equation}
f^{op}(a,G,M)=k\left( 1-P_{op}\right) .
\end{equation}%
Again to obtain the per capita\ average fertility payoff we multiply this by
the fraction of active individuals in the population $\sum_{j=1}^{w}G_{j}%
\left[ S_{j}^{m}+S_{j}^{f}\right] =\bar{S}^{m}+\bar{S}^{f}$. This leads to 
\begin{equation}
\bar{f}(a,G,M)=k\left( 1-P_{op}\right) \left( \bar{S}^{m}+\bar{S}^{f}\right)
=k\bar{S}^{f}.
\end{equation}

\section{ Appendix G Derivation of the age structured replicator equations}

The bracket describing the fertility stage of the gene pool dynamics will be%
\begin{gather}
\left( f_{g}(P_{j},a,G,M)-\bar{f}(a,G,M)\right) =\dfrac{k}{2}\left(
S_{j}^{f}+S_{j}^{m}\dfrac{1-P_{op}}{P_{op}}\right) -k\bar{S}^{f}= \\
\dfrac{k}{2}\left[ S_{j}^{f}+S_{j}^{m}\dfrac{\bar{S}^{f}}{\bar{S}^{m}}-2\bar{%
S}^{f}\right] =\dfrac{k}{2}\left[ S_{j}^{f}+\left[ \dfrac{S_{j}^{m}}{\bar{S}%
^{m}}-2\right] \bar{S}^{f}\right] = \\
k\left[ \dfrac{1}{2}\left( \dfrac{S_{j}^{f}}{\bar{S}^{f}}+\dfrac{S_{j}^{m}}{%
\bar{S}^{m}}\right) -1\right] \bar{S}^{f}.
\end{gather}%
The bracket describing the mortality stage will be 
\begin{gather}
\left( \bar{s}_{j}-\bar{s}\right) =\left( \sum_{i=0}^{m}a_{j}^{i}\bar{s}%
_{j}^{i}-\sum_{z=1}^{w}G_{z}\sum_{i=0}^{m}a_{z}^{i}\bar{s}_{z}^{i}\right) =
\\
\left( \sum_{i=0}^{m}a_{j}^{i}\left[ M_{j}^{i}s_{m}^{i}+\left(
1-M_{j}^{i}\right) s_{f}^{i}\right] -\sum_{z=1}^{w}G_{z}%
\sum_{i=0}^{m}a_{z}^{i}\left[ M_{z}^{i}s_{m}^{i}+\left( 1-M_{z}^{i}\right)
s_{f}^{i}\right] \right) ,
\end{gather}%
since $\bar{s}_{j}^{i}=M_{j}^{i}s_{m}^{i}+\left( 1-M_{j}^{i}\right)
s_{f}^{i} $ is the average survival of the $j$th strategy carrier in age
class $i$. Futhermore 
\begin{equation}
\bar{s}_{j}=\sum_{i=0}^{m}a_{j}^{i}\left[ M_{j}^{i}s_{m}^{i}+\left(
1-M_{j}^{i}\right) s_{f}^{i}\right] \text{ \ \ \ \ and \ \ \ \ \ }\bar{s}%
=\sum_{j=1}^{w}G_{j}\bar{s}_{j}
\end{equation}%
are the average $P_{j}$ carrier and population survival probabilities,
leading to 
\begin{equation}
\dot{G}_{j}=G_{j}\left( k\left[ \dfrac{1}{2}\left( \dfrac{S_{j}^{f}}{\bar{S}%
^{f}}+\dfrac{S_{j}^{m}}{\bar{S}^{m}}\right) -1\right] \bar{S}^{f}\left( 1-%
\dfrac{n}{K}\right) +\left( \bar{s}_{j}-\bar{s}\right) \right)
\end{equation}%
which is equation (\ref{srg}). Now we derive the equations describing the
dynamics of sex ratios in the particular age classes in the subpopulation of
carriers of strategy $P_{j}$ which will be equivalent to the original
equations on $M_{j}$. We use the version of equation (\ref{simsysb1}) for
two types (sexes) where the same payoffs are obtained in certain age ranges
specific for each type (thus the term $\sum_{i=0}^{m}a^{i}p_{j}^{i}f_{j}^{i}$
will be $f_{1}\sum_{i=a}^{b}a^{i}p_{1}^{i}$). In effect (\ref{simsysb1}) for
type $1$ (males)\ reduces to 
\begin{equation}
\dot{p}_{1}^{0}=\dfrac{f_{1}\sum_{i=a}^{b}a^{i}p_{1}^{i}-p_{1}^{0}\bar{f}}{%
a^{0}}\left( 1-\dfrac{n}{K}\right) ,
\end{equation}%
since $\bar{f}=\sum_{i=0}^{m}a^{i}\sum_{z=1}^{w}p_{z}^{i}f_{z}^{i}$ in (\ref%
{simsysb1}) is the average fertility in the population (in our case the
subpopulation of carriers of the $j$-th strategy). To translate the above
equation into the notation used in the sex ratio model we should apply the
following substations: $p_{1}^{0}\rightarrow M_{j}^{0}$, $f_{1}\rightarrow
f_{m}^{op}$, $f_{2}\rightarrow f_{f}^{op}$ and $\bar{f}\rightarrow f_{g}$
since $\sum_{i=a}^{b}a^{i}p_{1}^{i}$ is equivalent to $S_{j}^{m}$. Here
there are no different strategies indexed by a lower index but two opposite
sexes, thus for any particular gene we will have a single equation
describing the sex ratio in the zero age class: 
\begin{equation}
\dot{M}_{j}^{0}=\dfrac{\left(
f_{m}^{op}(P_{j},a,G,M)S_{j}^{m}-M_{j}^{0}f_{g}(P_{j},a,G,M)\right) }{%
a_{j}^{0}}\left( 1-\dfrac{n}{K}\right) .
\end{equation}%
Recalling (\ref{opratios}) that $\left( 1-M_{j}^{op}\right)
/M_{j}^{op}=S_{j}^{f}/S_{j}^{m}$ and $\left( 1-P_{op}\right) /P_{op}=\bar{S}%
^{f}/\bar{S}^{m}$ leads to equation (\ref{srm0}), 
\begin{gather}
\dot{M}_{j}^{0}=\dfrac{\left( \dfrac{k}{2}\left( \dfrac{1-P_{op}}{P_{op}}%
\bar{P}_{pr}+\dfrac{1-M_{j}^{op}}{M_{j}^{op}}P_{j}\right) S_{j}^{m}-M_{j}^{0}%
\dfrac{k}{2}\left( S_{j}^{f}+S_{j}^{m}\dfrac{1-P_{op}}{P_{op}}\right)
\right) }{a_{j}^{0}}\left( 1-\dfrac{n}{K}\right)  \notag \\
=\dfrac{k}{2a_{j}^{0}}\left( S_{j}^{m}\dfrac{\bar{S}^{f}}{\bar{S}^{m}}\bar{P}%
_{pr}+S_{j}^{m}\dfrac{S_{j}^{f}}{S_{j}^{m}}%
P_{j}-M_{j}^{0}S_{j}^{f}-M_{j}^{0}S_{j}^{m}\dfrac{\bar{S}^{f}}{\bar{S}^{m}}%
\right) \left( 1-\dfrac{n}{K}\right)  \notag \\
=\dfrac{k}{2a_{j}^{0}}\left( S_{j}^{m}\left( \bar{P}_{pr}-M_{j}^{0}\right) 
\dfrac{\bar{S}^{f}}{\bar{S}^{m}}+S_{j}^{f}\left( P_{j}-M_{j}^{0}\right)
\right) \left( 1-\dfrac{n}{K}\right) .
\end{gather}%
Note that the above equation is equivalent to the Tug of War dynamics of the
original model, and so should be completed by the respective equations for
all age classes (\ref{simsysb2}), which are equations (\ref{srm}), 
\begin{equation}
\dot{M}_{j}^{i}=\dfrac{a_{j}^{i-1}}{a_{j}^{i}}\left(
M_{j}^{i-1}s_{m}^{i-1}-M_{j}^{i}\bar{s}_{j}^{i-1}\right) .
\end{equation}%
The equations describing the age structure (\ref{simsysb3}) of the entire
population of carriers of the $j$-th strategy (with $f_{g}$ acting as $\bar{f%
}_{j}$) will be equations (\ref{sra})%
\begin{gather}
\dot{a}_{j}^{i}=a_{j}^{i-1}\bar{s}_{j}^{i-1}-a_{j}^{i}\left[
f_{g}(P_{j},a,G,M)\left( 1-\dfrac{n}{K}\right) +\bar{s}_{j}\right] =  \notag
\\
a_{j}^{i-1}\left[ M_{j}^{i-1}s_{m}^{i-1}+\left( 1-M_{j}^{i-1}\right)
s_{f}^{i-1}\right] -a_{j}^{i}\left[ \dfrac{k}{2}\left( S_{j}^{f}+S_{j}^{m}%
\dfrac{\bar{S}^{f}}{\bar{S}^{m}}\right) \left( 1-\dfrac{n}{K}\right) +\bar{s}%
_{j}\right] =  \notag \\
a_{j}^{i-1}\bar{s}_{j}^{i-1}-a_{j}^{i}\left[ \dfrac{k}{2}\left(
S_{j}^{f}+S_{j}^{m}\dfrac{\bar{S}^{f}}{\bar{S}^{m}}\right) \left( 1-\dfrac{n%
}{K}\right) +\bar{s}_{j}\right] .
\end{gather}%
From (\ref{N}) and (\ref{SRfaver}) we obtain the population size equation (%
\ref{srn}). Therefore we have derived the system (\ref{srg},\ref{sra},\ref%
{srm0},\ref{srm},\ref{srn}).

\begin{table}[tbp]
\begin{tabular}{l}
\hline
$n$ -population size \\ 
$n_{j}$ -number of individuals carrying the $j$-th strategy \\ 
$\tau $ -interaction rate \\ 
$f_{j}$ ($f_{j}^{i}$) -fertility payoff of the $j$-th strategy (of the $j$%
-th strategy at age $i$ ) \\ 
$s_{j}$ ($s_{j}^{i}=1-\tau d_{j}^{i}$) -survival payoff of the $j$-th
strategy (of the $j$-th strategy at age $i$ ) \\ 
$\bar{f}_{j}$, $\bar{f}^{i}$, $\bar{f}$ -average fertility for the $j$-th
strategy, $i$-th age class, whole population \\ 
$\bar{s}_{j}$, $\bar{s}^{i}$, $\bar{s}$-average survival for $j$-th
strategy, $i$-th age class, whole population \\ 
$\bar{r}=\tau \tilde{r}$ -Malthusian parameter, product of the interaction
rate and the game payoffs \\ 
$m+1$ -number of age classes \\ 
$w$ -number of strategies \\ 
$K$ -carrying capacity, maximal population load \\ 
$a_{j}^{i}$ -frequency of individuals at age $i$ among $j$-strategists \\ 
$p_{j}$ -frequency of $j$-strategists in the population \\ 
$a^{i}$ -proportion of individuals in the $i$-th age class \\ 
$p_{j}^{i}$ -frequency of $j$-strategists in the $i$-th age class \\ 
$P_{j}$ -sex ratio strategy (fraction of males in the brood of the female)
\\ 
$x$ ($x_{j}$) -number of females (carrying the $j$-th strategy) \\ 
$y$ ($y_{j}$) -number of males (carrying the $j$-th strategy) \\ 
$G_{j}=\left( x_{j}+y_{j}\right) /\sum_{l=1}^{w}\left( x_{l}+y_{l}\right) $
-frequency of the $i$-th strategy gene \\ 
$P=y/(x+y)$ -secondary sex ratio (proportion of males) \\ 
$\bar{P}_{pr}$ -primary sex ratio (average strategy of females) \\ 
$S_{j}^{f}=\sum_{l=c}^{d}a_{j}^{l}(1-M_{j}^{l})$ -proportion of active
females among the $j$-th strategy carriers \\ 
$S_{j}^{m}=\sum_{l=a}^{b}a_{j}^{l}M_{j}^{l}$ -proportion of active males
among the $j$-th strategy carriers \\ 
$M_{j}$ ($M_{j}^{i}$) -sex ratio of the population of the $j$th strategy
carriers (of $j$th strategy carriers at age $i$) \\ 
$k$ -number of offspring in the brood of a female \\ 
$M_{j}^{op}=S_{j}^{m}/\left( S_{j}^{m}+S_{j}^{f}\right) $ -operational sex
ratio of $j$-th strategists \\ 
$P_{op}=\bar{S}^{m}/\left( \bar{S}^{m}+\bar{S}^{f}\right) $ -operational sex
ratio in the population \\ \hline
\end{tabular}%
\caption{List of important symbols}
\end{table}

\textbf{Figure captions}

\textbf{Fig.1.} Schematic presentation of the discretization of the
continuous age dynamics. The assumed unit time step between age classes is
associated with a change of the population state, which may induce change of
the frequency dependent payoffs. However, while the resulting changes of the
vital rates are negligible, values of payoffs can be approximated by their
initial values at the beginning of the transition between age classes.

\textbf{Fig. 2.} The difference between two alternative formulations of the
problem: system (a) describes the evolution of the gene pool according to
age structures of carrier subpopulations, system (b) describes the evolution
of the global age structure driven by strategy selection in age classes.

\textbf{Fig 3.} The extension of the phase space of the sex ratio model to
the age structured case. The gene pool phase space is completed by
respective subspaces describing the age structures among carriers of the
particular genes, as in system $S_{a}$. Then each age structure subspace is
completed by subspaces describing carriers' sex ratios, according to system $%
S_{b}$.

\textbf{Fig.4. }Panel a) dynamics of gene frequencies, panel b) operational
sex ratios, for strategies $M_{0.05}^{op}$, $M_{0.55}^{op}$, $M_{0.95}^{op}$
and primary and operational sex ratios of the population $\bar{P}_{pr}$ and $%
P_{op}$, panel c) population size. Trajectories show that $P_{op}$ is the
threshold between growth and decline of the gene frequency depending on the
value of $M_{j}^{op}$. This is shown by the example of strategy $0.05$,
where bumps in the marked areas are caused by two types of events. The first
is when the strategy's operational sex ratio $M_{j}^{op}$ passes the
population's operational sex ratio $P_{op}$, which is the threshold between
growth and decline. The second is when the average operational sex ratio $%
P_{op}$ passes the value of 0.5 which inverts the strategic situation, since
the opposite sex is in the minority when this happens.

\textbf{Fig.5.} Trajectories of age classes. The initial behaviour is caused
by huge differences in the initial sex ratios. The assumed changes in age
specific survivals slightly affect the trajectories.

\textbf{Fig.6.} Trajectories of age specific sex ratios. The pattern caused
by the assumed changes in survival probabilities is clearly visible.

\textbf{Fig.7. } A plot of the convergence to the respective Euler-Lotka
manifolds (dashed lines) for arbitrarily chosen age classes for strategy $%
0.05$. The convergence is delayed by some inertia caused by the age dynamics.

\textbf{Fig.8}. Plots of the excess fertility payoffs $\left(
f_{g}(P_{j},a,G,M)-\bar{f}(a,G,M)\right) $, excess mortality payoffs $\left( 
\bar{s}_{j}-\bar{s}\right) $ and the gene frequency growth rates \newline
$\left( f_{g}-\bar{f}\right) \left( 1-n/K\right) +\left( \bar{s}_{j}-\bar{s}%
\right) $ from the gene pool dynamics (\ref{srg}). Fertility payoffs are not
equal as in the classical theory, and the same situation is true for
mortality payoffs, but the right hand sides of the equations are zero. This
shows that the explanation for the primary sex ratio being 0.5 needs an
explicit consideration of the interplay between fertility and mortality.

\begin{figure}[h]
\includegraphics[width=15cm]{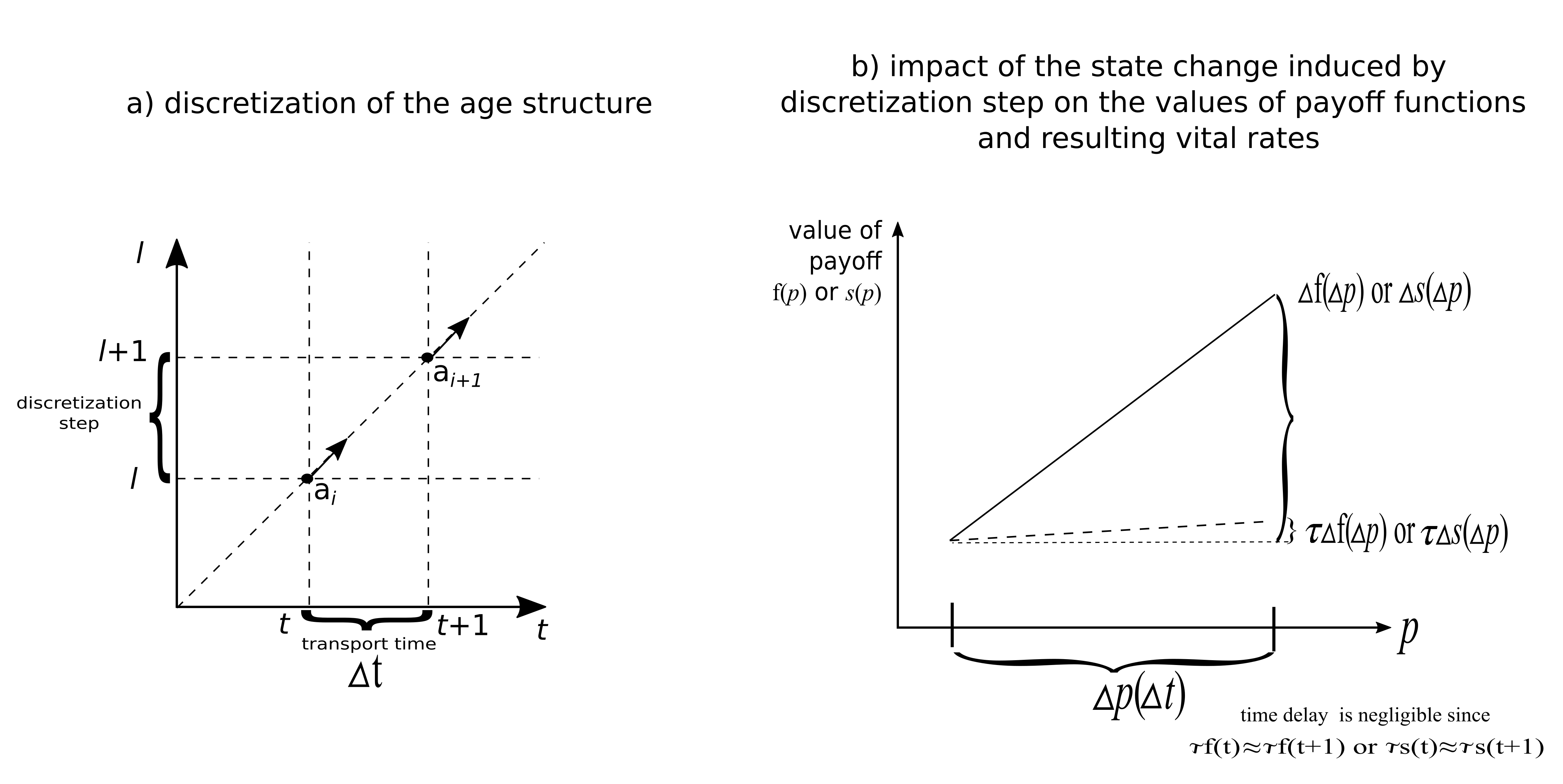} \centering
\end{figure}

\begin{figure}[h!]
\includegraphics[width=15cm]{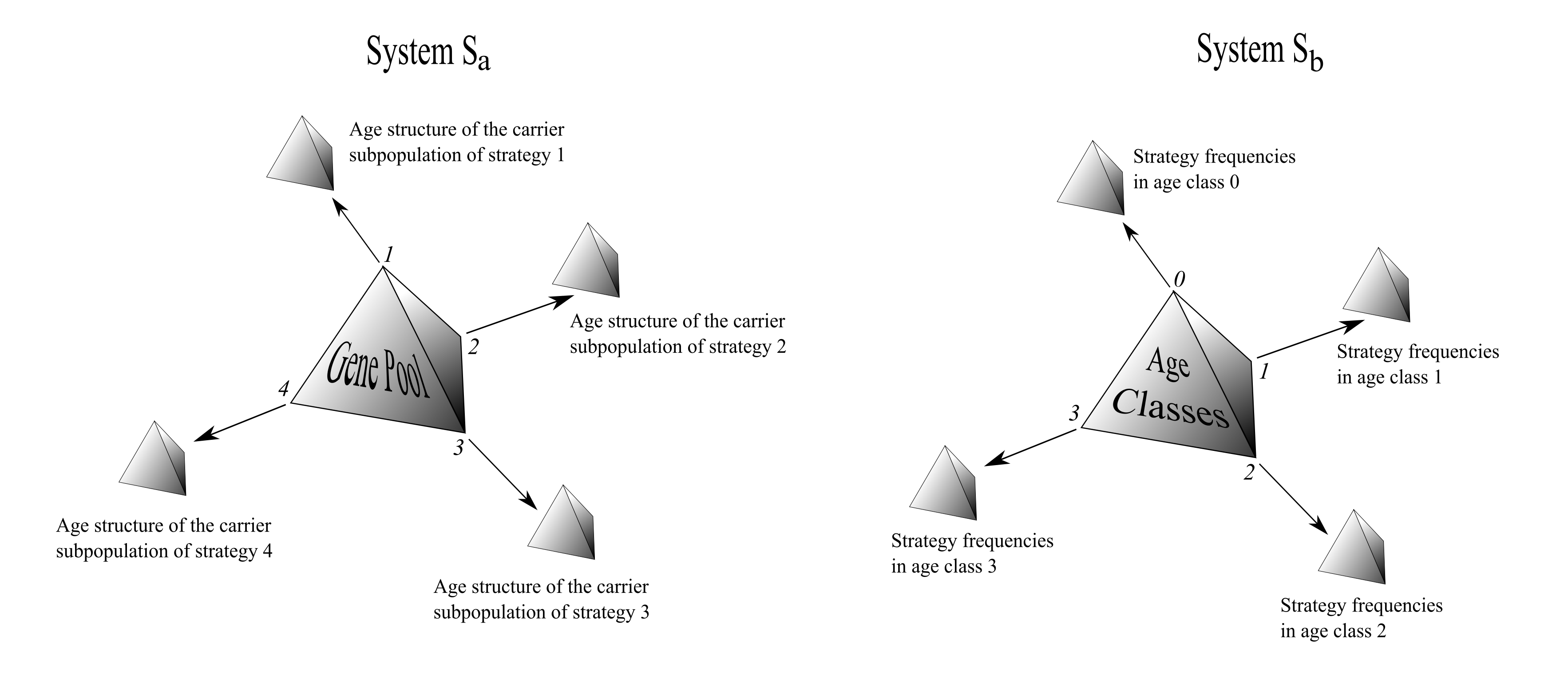} \centering
\end{figure}

\begin{figure}[h!]
\includegraphics[width=15cm]{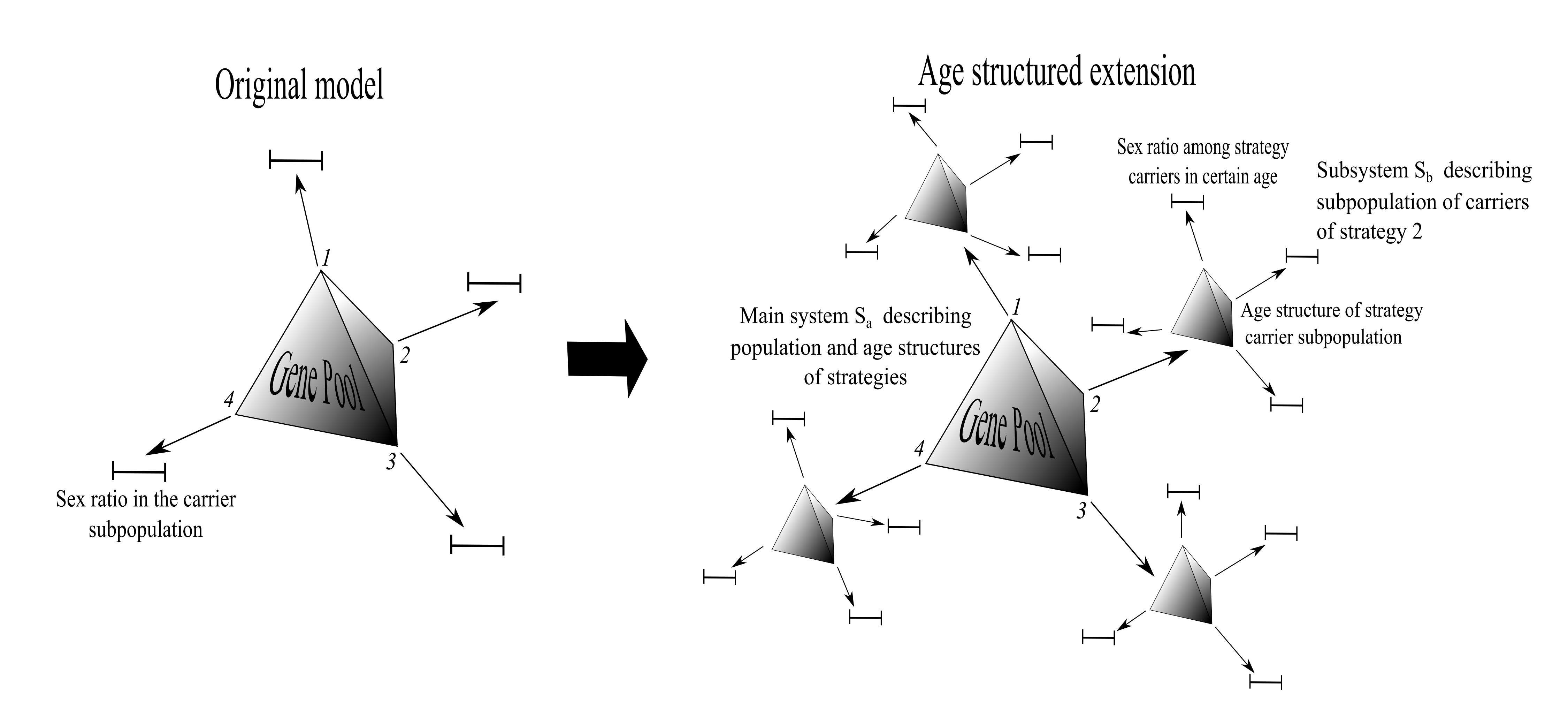} \centering
\end{figure}

\begin{figure}[h!]
\includegraphics[width=15cm]{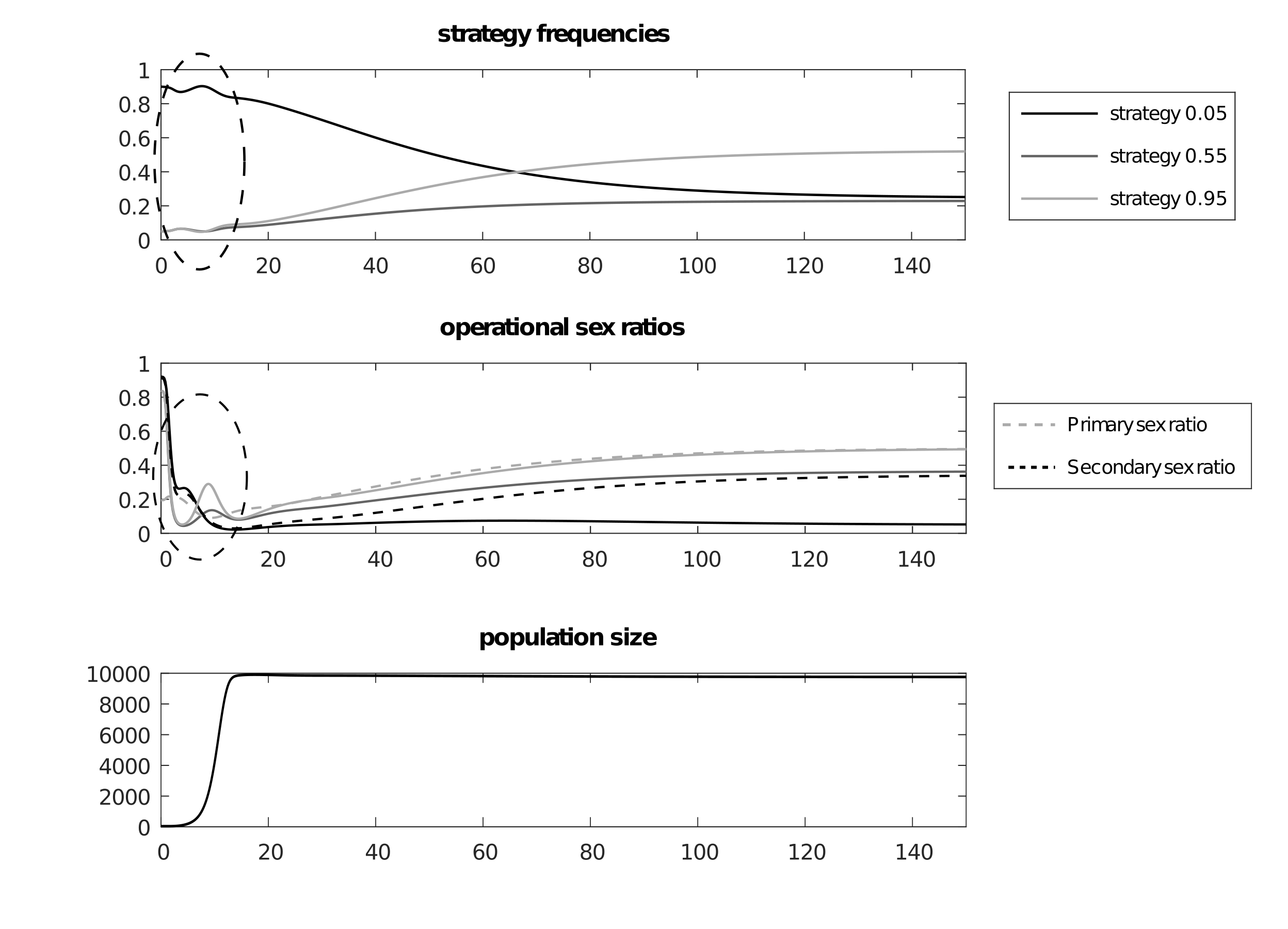} \centering
\end{figure}

\begin{figure}[h!]
\includegraphics[width=15cm]{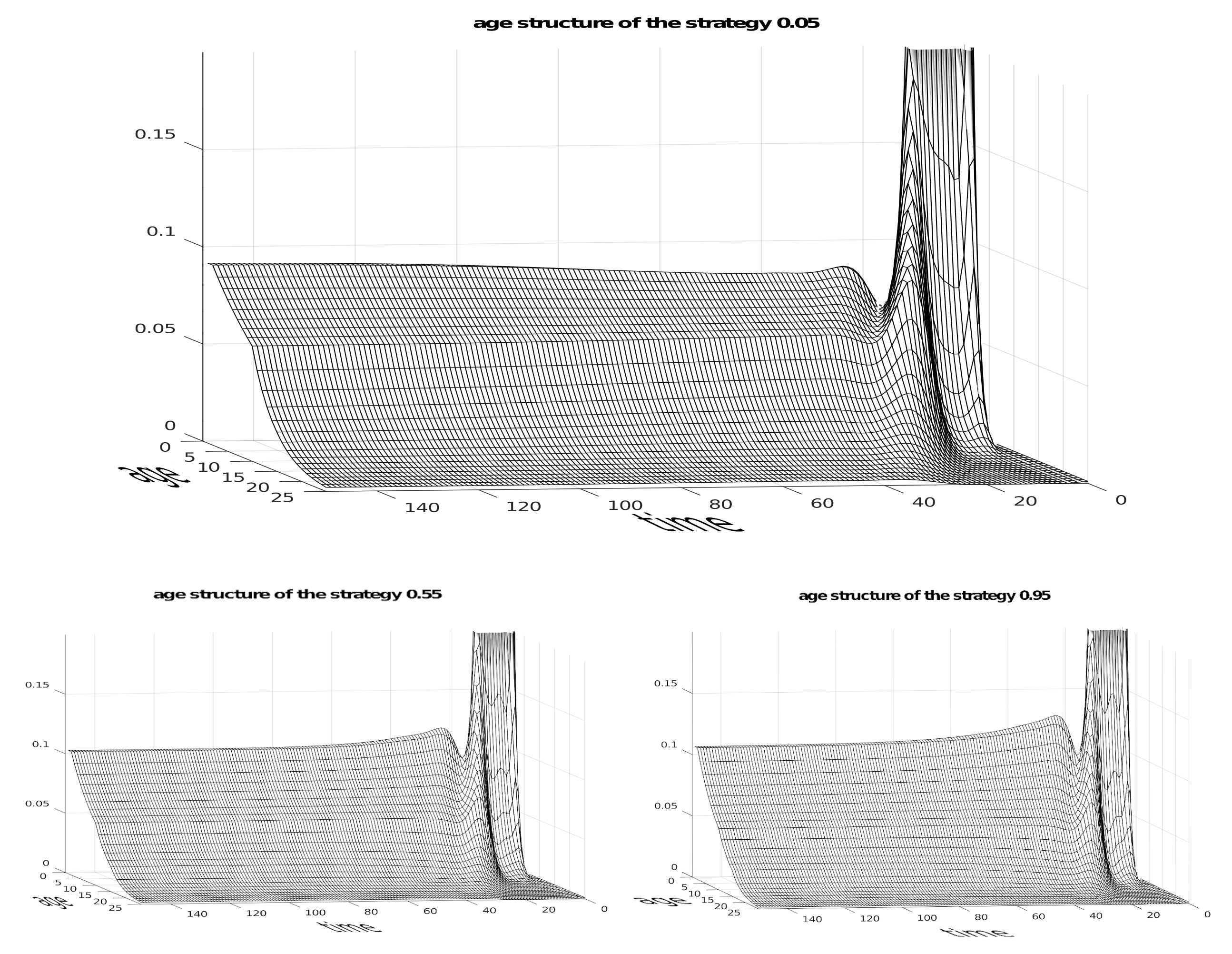} \centering
\end{figure}

\begin{figure}[h!]
\includegraphics[width=15cm]{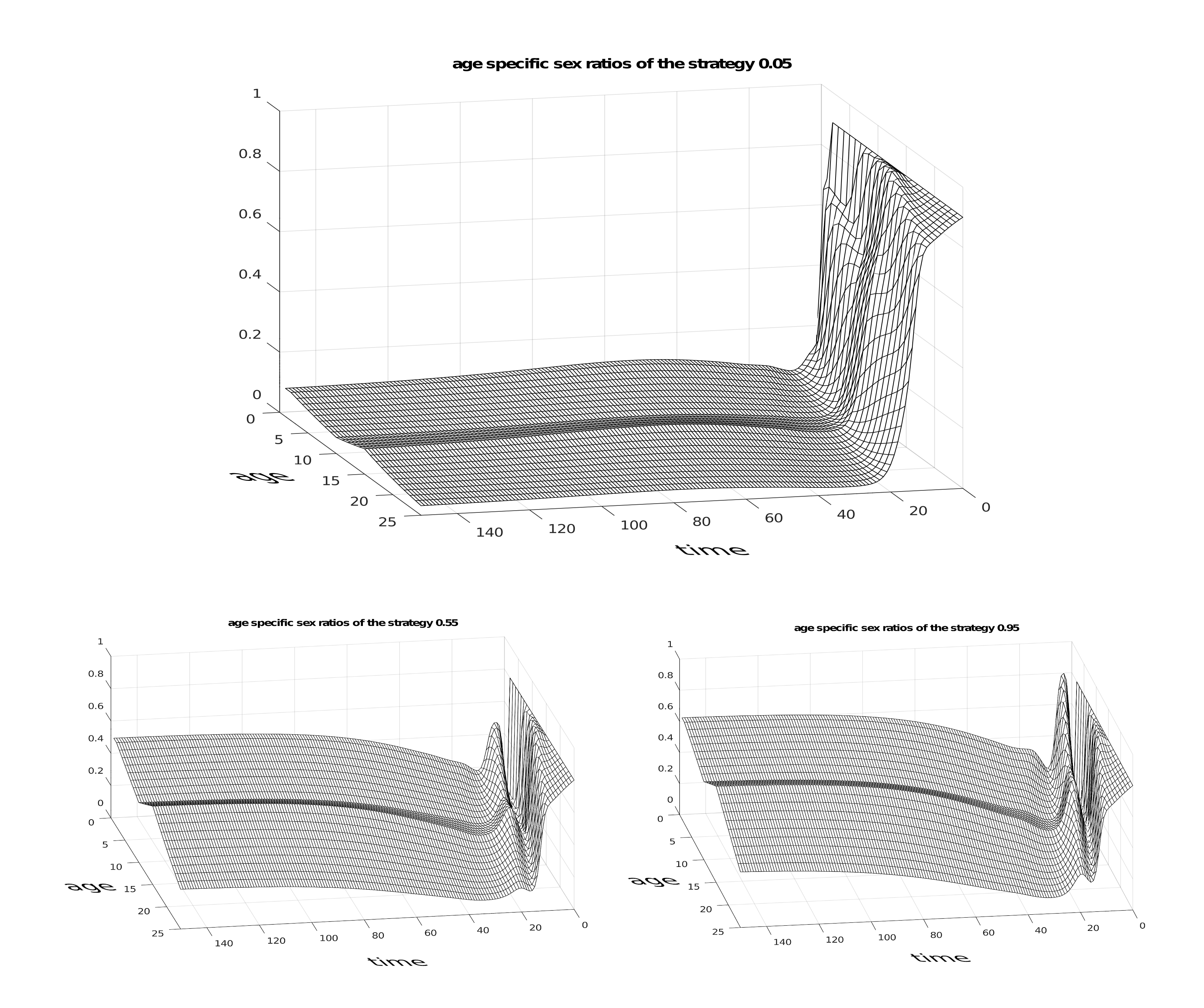} \centering
\end{figure}

\begin{figure}[h!]
\includegraphics[width=15cm]{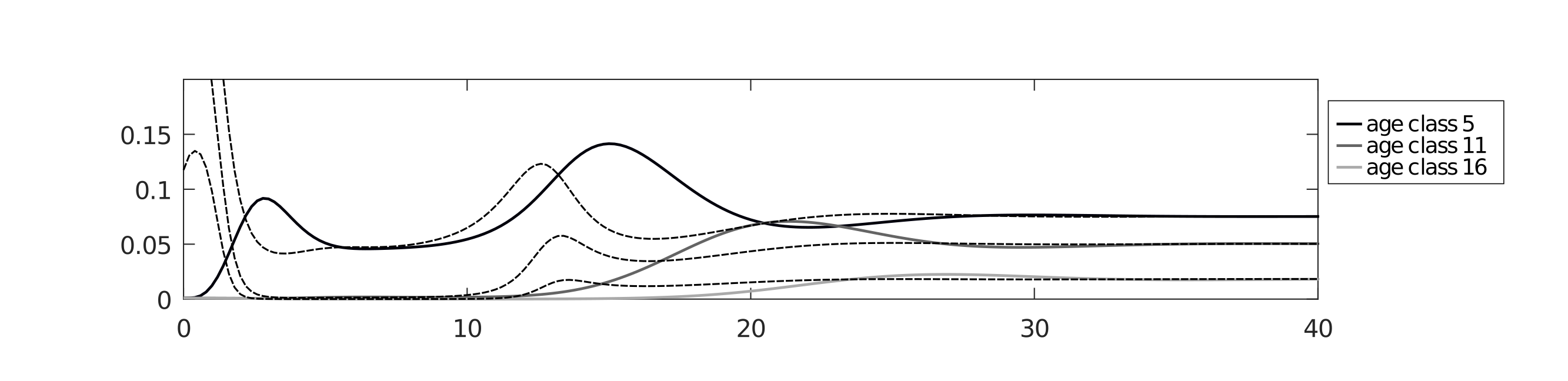} \centering
\end{figure}

\begin{figure}[h]
\includegraphics[width=15cm]{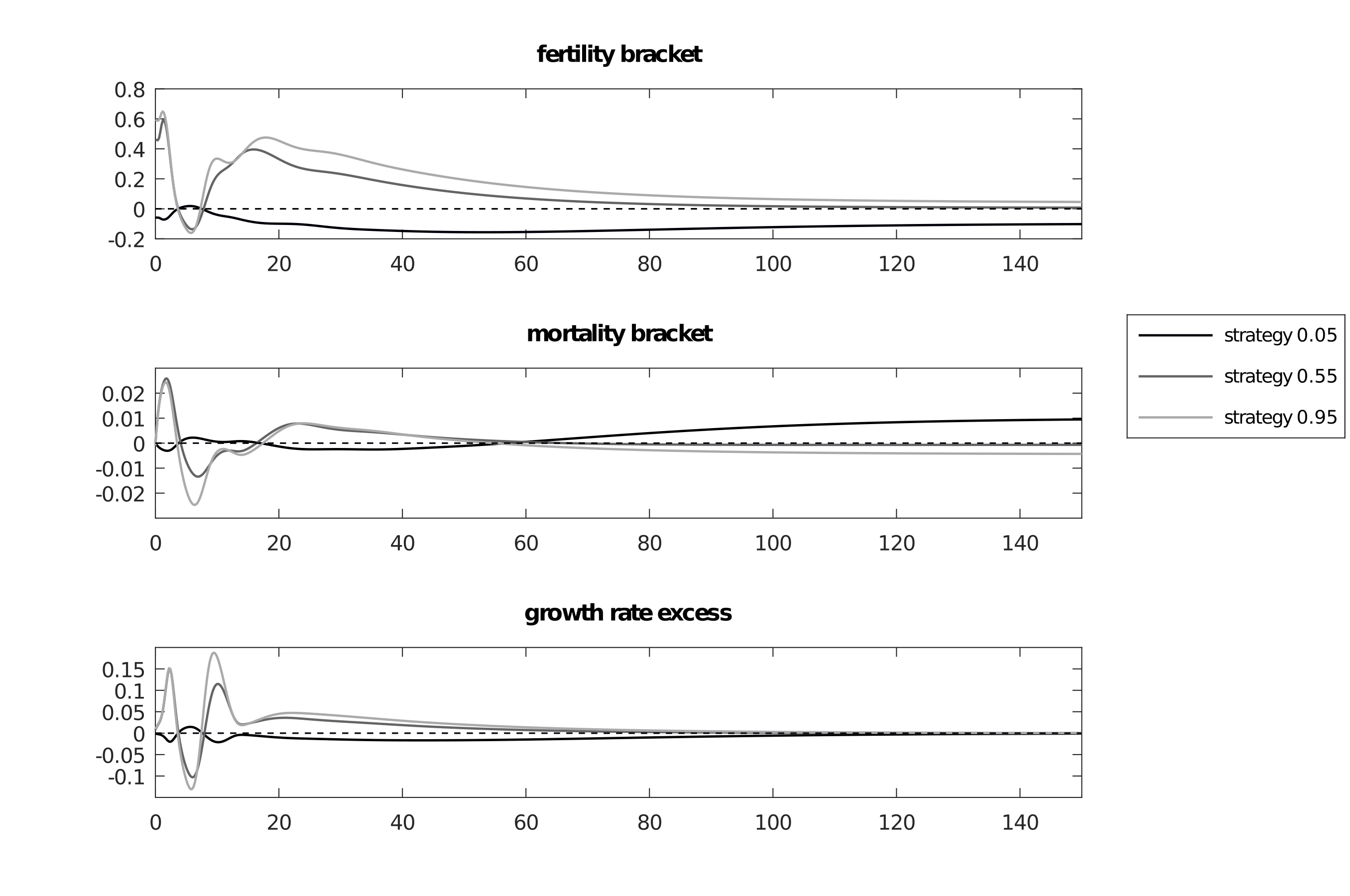} \centering
\end{figure}

\end{document}